\documentclass[]{jfm}   

\usepackage{xcolor}

\newcommand{\ind}{\mathbb{I}}

\usepackage{graphicx}
\usepackage{amsmath}
\usepackage{bm}
\usepackage{newtxtext}
\usepackage{newtxmath}
\usepackage{natbib}
\usepackage{hyperref}
\hypersetup{
    colorlinks = true,
    urlcolor   = blue,
    citecolor  = black,
}

\newcommand{\RomanNumeralCaps}[1]
\linenumbers

\newcommand{\JP}[1]{{\color{black}  #1}}
\newcommand\REV[1]{{\color{black}#1}}

\title{ Dynamical relevance of periodic orbits under increasing Reynolds number and connections to inviscid dynamics} 

\author{Andrew Cleary\aff{1},
  Jacob Page\aff{1}\corresp{\email{jacob.page@ed.ac.uk}}}

\affiliation{\aff{1} School of Mathematics and Maxwell Institute for Mathematical Sciences, University of Edinburgh, Edinburgh, EH9 3FD, UK }

\begin{document}

\maketitle

\begin{abstract}
Large numbers of \REV{relative periodic orbits (RPOs)} have been found recently in doubly-periodic, two-dimensional Kolmogorov flow at moderate \JP{Reynolds numbers} $\Rey \in \{40, 100\}$.
While these solutions lead to robust statistical reconstructions at the $\Rey$-values where they were obtained, it is unclear how their dynamical importance evolves with increasing $\Rey$.
\JP{
We perform arclength continuation on this library of solutions to show that large numbers of RPOs quickly become dynamically irrelevant, reaching dissipation values either well above or below those associated with the turbulent attractor at high $Re$. 
The scaling of the high dissipation RPOs is shown to be consistent with a direct connection to solutions of the unforced Euler equation, and is observed for a wide variety of states beyond the `unimodal’ solutions considered in previous work (Kim \& Okamoto, \emph{Nonlinearity} \textbf{28}, 2015). 
On the other hand, the weakly dissipative states have properties indicating a connection to exact solutions of a forced Euler equation. 
The apparent dynamical irrelevance is associated with poor statistical reconstructions away from the $Re$ values where the RPOs were originally converged. 
Motivated by the connection to solutions of the Euler equation, we show that many of these states can be well described by exact relative periodic solutions in a system of point vortices. 
}
The point vortex RPOs are converged via gradient-based optimisation of a scalar loss function which (1) matches the dynamics of the point vortices to the turbulent vortex cores and (2) insists the point vortex evolution is itself time-periodic.
\end{abstract}


\section{Introduction}
\label{sec:introduction}

The dynamical systems view of turbulence considers turbulent flows as trajectories in a high dimensional space, pin-balling between simple invariant solutions, such as equilibria, travelling waves (TWs) or unstable periodic orbits (UPOs) \citep{Kawahara2012, Graham2021}. 
This perspective gained considerable interest in recent years following the discovery of unstable travelling waves on the laminar-turbulent boundary in a pipe \citep{Faisst2003, wedin_kerswell_2004} and a \REV{UPO} in a minimal Couette flow \citep{Kawahara2001}.
Since then there have been systematic searches for these simple invariant solutions in a myriad of flow configurations including parallel pipes \citep{Kerswell2005,Budanur2017}, channels \citep{Waleffe2001,Park_Graham_2015}, plane Couette flow \citep{Gibson2008, Krygier2021}, periodic, body-forced two-dimensional \citep{Chandler2013, lucas2015, suri2020, page2022recurrent, Zhigunov2023} and three-dimensional flows \citep{yalniz2021, lucas2017}, to list only a few.

There have been promising instantaneous observations of apparent visits to UPOs in experiments and simulations \citep{hof2004,suri2017,suri2018}, with perhaps the most convincing being the Taylor-Couette study from \citet{Krygier2021} showing clear evidence of \REV{ shadowing of relative periodic orbits (RPOs)} by the turbulence over multiple periods.
This observed shadowing is consistent with the view in periodic orbit theory, which was developed to predict statistics of chaotic attractors in uniformly hyperbolic dynamical systems \citep{ChaosBook, Artuso1990a, Artuso_1990b, CVITANOVIC1991}.
Periodic orbit theory expresses statistics as weighted sums of the statistics of the UPOs, where
the weights are determined by the local stability of each solution \citep{Christiansen1997}.
\JP{However, rigorous application of periodic orbit theory requires a complete library of solutions \citep[see discussion in][]{cvitanovic2013}, and attempts to use the formulae on a modest set of RPOs in two-dimensional turbulence proved no better than simply assigning an equal weight to each solution \cite{Chandler2013,lucas2015}.}
This has motivated recent data-driven approaches for `optimal' weight estimation within an incomplete library of periodic orbits \citep{yalniz2021,page2022recurrent,Redfern2024}. 

A long-standing limitation to these statistical predictions has been the computation of sufficiently large libraries of dynamically important simple invariant solutions.
Recent attempts to remove reliance on classical recurrent flow analysis to detect UPOs have had some success in significantly increasing the number of converged solutions, leading to relatively robust statistical coverage with periodic orbits in moderate-$\Rey$ Kolmogorov flow \citep{page2022recurrent,Redfern2024}.
The hope has been that, with good statistical coverage at one value of $\Rey$, predictions can then be made at higher Reynolds numbers by tracking the UPOs along their solution branches, though at present
we have only limited understanding of the range of $\Rey$ over which we can expect robust reproduction of turbulent statistics.
Previous continuation efforts have shown that solutions often \JP{become increasingly dynamically irrelevant} as $\Rey$ is increased, with many moving to the laminar-turbulent boundary in bistable flows \citep{waleffe2003, wang2007}.
\JP{Continuation in wall-bounded flows also has to contend with the emergence of multiscale structure not contained in the low-$Re$ solutions, and efforts have focused on looking for solutions with this property in reduced order models \citep{McCormack2024}.}
Furthermore, solutions often emerge in saddle node bifurcations, while the solution branches themselves may be highly non-monotonic in $\Rey$ \citep{Gibson2008,Chandler2013}.
An additional layer of complexity is the need to also understand how the weights in the statistical expansions should change under increasing Reynolds number.
Here we make a data-driven assessment of both (i) the range of statistical coverage (in $\Rey$) that is obtainable with a set of solutions converged at a single parameter setting and (ii) the dependence of the statistical weightings as $\Rey$ is adjusted.

A large library of converged UPOs will contain a variety of recurrent processes.
Systematically identifying the relevant physical phenomena is a significant challenge. 
In three-dimensional shear flows, asymptotic theories have emerged to describe roll/streak interactions \citep{Hall_Smith_1991,hall_sherwin_2010} and the `staircase’-like structure of the instantaneous streamwise velocity \citep{Montemuro2020}.
\JP{In two-dimensional turbulence there is evidence that `dynamically relevant' UPOs connect directly to exact solutions of the inviscid Euler equation, which can simplify the elucidation of the underlying mechanics.}
This connection has been conjectured for a particular class of `unimodal' solutions in two-dimensional Kolmogorov flow \citep{Kim2015,Kim2017} which share similar vortical features with the condensate that dominates undamped two-dimensional turbulence at high $\Rey$ \citep{smith1993}, and which become insensitive to the background forcing required to maintain them in the presence of viscosity.
A collection of exact solutions to the Euler equation (with a regularising hyperviscous term) was obtained recently by \citet{Zhigunov2023}. 
The authors were able to show that many of the exact Euler solutions \JP{dominated by a large vortex pair} were relevant to the dynamics of finite-$Re$ turbulence.

A generic feature of the various Euler solutions reported in \citet{Zhigunov2023} is that they exist as members of infinite-dimensional continuous families.
This feature is observed in other analytical (smooth) solutions to the Euler equations, with perhaps the most well-known being
`Stuart vortices' \citep{stuart1967}, 
a row of spanwise `cat's eye' vortices with uniform counter-flowing streams in the far field.
The continuous family depends on a single parameter, which smoothly deforms the flow from a parallel $\tanh$ mixing layer through the cat's-eye structures to a row of point vortices.
Analogous solution families have been obtained on the surface of a stationary sphere \citep{Crowdy2003a} and the 2-torus \citep{sakajo2019}.
Other inviscid solutions with some relevance to the various Navier-Stokes UPOs found in the present study include multipolar solutions consisting of a central core vortex surrounded by a number of opposite-signed satellite vortices. 
These structures have been found analytically as superpositions of vortex patches with line vortices \citep{Crowdy1999} and in numerical simulations of instability growth on axisymmetric vortices of various forms \citep{carton1989,Morel1994,Carnevale1994}.

Connections to exact solutions of the inviscid equations have also been conjectured in decaying two-dimensional turbulence.
For instance, \citet{jimenez_2020} showed that most of the kinetic energy in the early stages of decay, during which the dominant scale of the kinetic energy is small compared to the domain, is contained in a slowly evolving quasi-equilibrium `crystal' of vortex cores. 
The significance of point-vortex dynamics in the description of two-dimensional turbulence is well-known, with compelling numerical evidence that \JP{the dynamics of finite-area vortices can be described by a `punctuated Hamiltonian' model \citep{benzi1992}.}
\JP{In this model, vortex motion is governed by the equations of point vortex dynamics, interrupted only by non-Hamiltonian merger events.}
There are large numbers of vortex crystals \citep[this terminology is reserved for exact relative equilibria with fixed inter-vortex distances, see the review][]{Aref2003} documented in the literature for various numbers of vortex cores, $N_v$. 
Many solutions have been obtained via geometrical or symmetry considerations \citep[e.g.][]{stieltjes1900, lewis1996, aref2005}. 
In another approach, free-energy minimisation has been employed to find extremely large numbers of crystals for large $N_v = O(50)$ \citep{campbell1979vortex,Cleary2023}.
The latter study made use of gradient based optimisation on augmented loss functions to search for crystals with specific features, and is adapted here to find exact point vortex solutions that replicate the key vortex dynamics contained in a library of Navier-Stokes UPOs.

The aims of this paper are twofold. 
The first is to investigate how the dynamical importance of RPOs in two-dimensional Kolmogorov flow \JP{changes with increasing} $\Rey$. 
This is achieved by performing a large arclength continuation in $Re$ of a library of RPOs, and tracking their contribution to the turbulent statistics. 
\JP{One intriguing feature of the solutions obtained at relatively low $Re$ is that many appear to connect directly to solutions of the unforced Euler equation as $Re\to \infty$.}
\JP{As a result, a second aim of this manuscript is to attempt to} represent the dynamics of the self-sustaining, large-scale coherent vortices in the RPOs with exact solutions of the point vortex system. 
In section \ref{sec:comp_setup}, we formulate the equations of motion of two-dimensional Kolmogorov flow and the point vortex system in doubly periodic domains. 
In section \ref{sec:forced_turb}, we present the large continuation effort of turbulent UPOs. 
In section \ref{sec:pv_matching}, we present the labelling of the turbulent UPOs via solutions of the point vortex model. 

\section{Formulation}
\label{sec:comp_setup}

We begin this section by introducing two-dimensional Kolmogorov flow and the associated library of RPOs (§\ref{sec:comp_kf}) which seed the continuation effort presented in \ref{sec:forced_turb}. 
We \JP{present a method to extract vortex cores from the RPOs in \S\ref{sec:vort_id}, before formulating the doubly-periodic point vortex system in \S \ref{sec:comp_pv}, with which we will model the large-scale vortex dynamics in the Kolmogorov solutions}.

\subsection{Kolmogorov Flow}
\label{sec:comp_kf}

We consider two-dimensional turbulence in a square domain with periodic boundary conditions, driven by a monochromatic body force in the streamwise direction. 
The out-of-plane vorticity $\omega = \partial_x v - \partial_y u$, where the velocity $\bm{u} = (u,v)$, evolves according to
\begin{equation}
    \partial_t \omega + \bm{u} \cdot \boldsymbol\nabla{\omega} = \frac{1}{\Rey} \nabla^2 \omega - n \cos {ny}.
    \label{eq:kf_eq}
\end{equation}
In this non-dimensionalisation we have chosen a length scale $1/k^{*}$ as the inverse of the fundamental wavenumber of the domain $k^* = 2\pi / L^*$, and a time scale $1/\sqrt{k^*\chi^*}$, where $\chi^*$ is the amplitude of the forcing in the momentum equation. 
These length and time scales lead to the definition of the Reynolds number in this flow $\Rey \coloneq \sqrt{\chi^*/k^{*3}}/\nu$, where $\nu$ is the kinematic viscosity.
Throughout this work, we set the forcing wavenumber $n = 4$, as has been common in previous studies \citep{Chandler2013, Page2021, page2022recurrent}.

Equation (\ref{eq:kf_eq}) is equivariant under continuous shifts in the streamwise ($x$) direction, $\mathscr{T}_s : \omega(x,y) \to \omega(x+s, y)$, under discrete shift-reflects by a half-wavelength in $y$, $\mathscr{S}: \omega(x,y) \to -\omega(-x, y+\pi/4)$ and under discrete rotations by $\pi$, $\mathscr R : \omega(x,y) \to \omega(-x,-y)$. 
\JP{The existence of the continuous symmetry means that we generically expect exact solutions to be relative periodic orbits which shift a finite distance in $x$ over one period.
This is the case for the vast majority of solutions in our library.
}

\REV{Some key integral observables reported throughout this work are the total kinetic energy
\begin{equation}
    E(t) \coloneq \frac{1}{2} \langle \bm{u}^2 \rangle_V,
\end{equation}
the total dissipation rate
\begin{equation}
    D \coloneq \frac{1}{\Rey} \langle | \nabla\bm{u} | ^2 \rangle_V = \frac{1}{\Rey} \langle \omega^2 \rangle_V,
\end{equation}
and the total production rate
\begin{equation}
    I \coloneq \langle u \sin(ny) \rangle_V,
\end{equation}
where the average over the volume $V$ is defined as
\begin{equation}
    \langle \bullet \rangle_V \coloneq \frac{1}{(2\pi)^2} \iint \bullet  \; d^2\bm{x}.
    \label{eq:spatial_av}
\end{equation}
}

We solve equation (\ref{eq:kf_eq}) using the spectral version of the \texttt{JAX-CFD} solver \citep{Kochkov2021, LCspectral}.
At each time-step in the solver, the velocity field is computed by solving the Poisson equation $\nabla^2 \psi = -\omega$, where the streamfunction $\psi$ is related to the induced velocity components via $u = \partial_y \psi, v = -\partial_x \psi$.
The resolution is varied depending on the value of $\Rey$: $N_x\times N_y = 128 \times 128$ when $\Rey \leq 150$, rising to $256\times 256$ when $150 < \Rey \leq 300$, and finally $512 \times 512$ when $\Rey >300$.

Our analysis begins with a large set of RPOs converged at $Re=40$ and $Re=100$, which are documented in \citet{Page2024} and \citet{page2022recurrent} (see those papers for full details of the solutions, including periods, shifts, Floquet exponents etc). 
\REV{Altogether,} we begin with 174 RPOs at $\Rey = 40$ and 151 RPOs at $\Rey = 100$. 
In order to investigate how the dynamical importance of these solutions varies with $\Rey$, we perform a continuation of the solutions at $\Rey = 40$ and $\Rey = 100$ to higher Reynolds number. 
The continuation results in a large set of solutions in the region $\Rey \in (30, 1100)$, with each RPO belonging to a solution branch initialised at either $\Rey = 40$ or $\Rey = 100$.
Branches can go through multiple \REV{fold} bifurcations throughout the continuation, so that $\Rey$ is not necessarily monotonic along the curves.

\subsection{Vortex identification}
\label{sec:vort_id}

\begin{figure}
    \centering
    \includegraphics[width=0.9\linewidth]{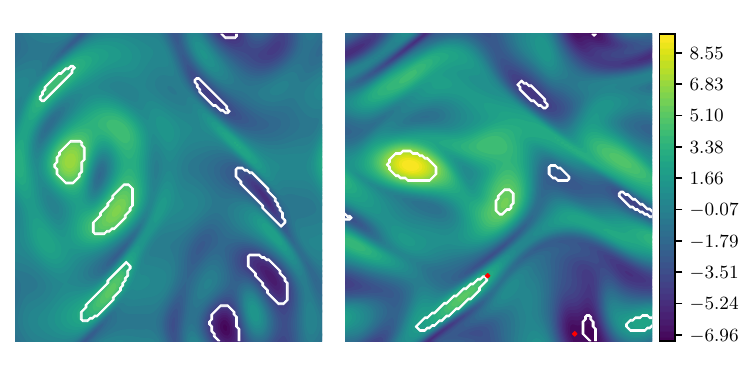}   
    \caption{Application of the vortex extraction criterion discussed in the text to two example snapshots at $Re=100$. Contours are of the out-of-plane vorticity, white lines identify regions identified by the threshold (\ref{eq:vort_extraction_cond}), and red lines \JP{(which here enclose areas so small they appear as points in the visualisation)} highlight which of these regions were discarded due to the bound imposed on the region area. }
    \label{fig:extraction_eg}
\end{figure}
\JP{Motivated by a possible connection of exact solutions in viscous Kolmogorov flow to solutions of the Euler equation as $Re \to \infty$ \citep{Kim2015,Kim2017,Zhigunov2023}, part of this work explores the ability of point vortex solutions to capture the dynamics of the large scale vortices in the Navier-Stokes RPOs.}
\JP{Note that while a Kolmogorov forcing profile is used throughout this work, previous work \citep{Kim2017, Gallet2013, Zhigunov2023} indicates that the large-scale structure of the flow may detach from the forcing in equation (\ref{eq:kf_eq}) at very high $\Rey$.}

In order to model the vortex cores of the Kolmogorov RPOs with point vortices, we must first identify and extract coherent vortices from the turbulent flows.
We follow the methodology in \citet{Page2024} and first compute the root-mean-square vorticity fluctuations $\omega_{RMS} \coloneq \sqrt{\langle (\omega(\bm{x},t) - \overline{\omega}(y))^2  \rangle}_V$, where the overline denotes an average over time, $x$ direction and ensemble. 
Spatially localised vortices are then extracted as any connected regions $\mathcal{V}_i$ where 
\begin{equation}
    | \omega(\bm{x},t) - \omega_{RMS} | \ge 2 \omega_{RMS}.
    \label{eq:vort_extraction_cond}
\end{equation}
Two examples of these connected regions extracted from a vorticity snapshot are shown in figure \ref{fig:extraction_eg}. 
Note that this method requires that a reference vorticity magnitude must be chosen ($2 \omega_{RMS}$ in (\ref{eq:vort_extraction_cond})), which influences the boundaries of the extracted vortex cores. 
This method is agnostic to the shape of the connected region, so that strong vorticity filaments are also occasionally extracted as vortex `cores' -- note the thin structure stretching across the periodic boundary in the right panel of figure \ref{fig:extraction_eg}. 

Following vortex boundary identification, the circulation $\Gamma_i \coloneq \int_{\bm{x} \in \mathcal{V}_i} \omega(\bm{x}) d\bm{x}$, vortex area $\mathcal{A}_i \coloneq \int_{\bm{x} \in \mathcal{V}_i} \ind(\bm{x}) d\bm{x}$ and the centre of vorticity $\tilde{\bm{x}}_{i} = (\tilde{x}_i, \tilde{y}_i) = \frac{1}{\Gamma_i} \int_{\bm{x} \in \mathcal{V}_i} \omega(\bm{x}) \big( x dx, y dy \big)$ of each vortex core can all be computed.
Similar to the approach in \citet{Benzi1987}, we consider only `large-scale' vortices, ignoring any structures with area less than 10\% of the largest vortex $\mathcal{A}_i < 0.1 \max_j(\mathcal{A}_j)$ (see discarded vortices in red in figure \ref{fig:extraction_eg}). 
Finally, to satisfy the Gauss constraint, $\int \omega(\bm{x})d\bm{x} = 0$, we must ensure that $\sum_j \Gamma_j = 0$, which is enforced by equally modifying the extracted circulation of each vortex core, $\Gamma_j \to \Gamma_j - \langle \Gamma \rangle$ prior to initialising a point vortex computation.

\JP{
Our attempt to fit point vortex solutions is performed at $Re=100$. 
At this value we find an average $\langle|\Gamma|\rangle = 2.84$ with standard deviation $\sigma_{\Gamma} = 2.68$ for the extracted cores in the RPO library \citep[the distribution has a long right tail, see][]{Page2024}. 
As a result, the induced velocities are $O(1)$.
This should be contrasted to the much weaker time average velocity, which is around an order of magnitude smaller (statistics are presented in \S\ref{sec:stats}). 
}

\subsection{The doubly periodic point vortex model}
\label{sec:comp_pv}
\JP{We now outline the formulation for the point vortex problem in a doubly periodic domain.}
The \REV{following} equations of motion \JP{in this system were first derived by} \citet{weiss1991}. 
An alternative derivation is presented here following the approach adopted by \citet{Aref2003} for a periodic strip.
We begin by considering $N_v$ point vortices in an infinite two-dimensional domain \JP{without a background velocity}, where the $j^{th}$ vortex is located at the complex position $z_{j} = x_{j} + iy_{j}$. 
Each point vortex moves under the induced velocity from the $N_v-1$ other vortices in the domain. 
The equations of motion for these point vortices are
\begin{equation}
    \dot{\overline{z}}_j = \dot{x}_j - i\dot{y}_j = \frac{1}{2\pi i} \sum_{k = 1}^{N_v}  \phantom{}^{'} \frac{\Gamma_k}{ z_j - z_k},
\end{equation}
where the overbar represents the complex conjugate, the dot denotes a time derivative, $\Gamma_k$ is the circulation of the $k^{th}$ point vortex and the prime indicates the omission of any singular terms (i.e. no self-induced velocity from $j = k$). 
These equations form a Hamiltonian system \cite[e.g. see][]{batchelor1967}, 
\begin{equation}
    \Gamma_j \dot{\overline{z}}_j = -\frac{1}{i}\frac{\partial H_u}{\partial z_j},
\end{equation}
where the unbounded Hamiltonian is
\begin{equation}
    H_u = -\frac{1}{4\pi}\sum_{j=1}^{N_v}\sum_{k=1}^{N_v}\phantom{}^{'} \Gamma_{j} \Gamma_{k} \log{|z_{j} - z_{k}|}.
    \label{eq:h_unbounded}
\end{equation}
This result can be generalised to the bounded, doubly periodic $L \times L$ domain by introducing ghost vortices at $z_j + nL + imL$ with $m,n \in \mathbb{Z} \backslash \{ 0 \}, j \in \{1, \dots, N_v \}$. 
The equations of motion then become
\begin{equation}
    \dot{\overline{z}}_j = \frac{1}{2\pi i} \sum_{n,m = -\infty}^{\infty} \sum_{k = 1}^{N_v} \phantom{}^{'} \frac{\Gamma_k}{z_j - z_k - nL - imL}.
    \label{eq:eom_initial}
\end{equation}
We can make use of the identity
\begin{equation}
    \sum_{n=-\infty}^{\infty} \frac{1}{x-n} = \pi \cot(\pi x)
    \label{eq:sum_identity}
\end{equation}
to simplify equation (\ref{eq:eom_initial}) by evaluating either the summation over $n$
\begin{equation}
    \dot{\overline{z}}_j = \frac{1}{2 Li} \sum_{k} \Gamma_k \sum_{m=-\infty}^{\infty} \cot\left(\pi\left(\frac{z_j-z_k}{L} - im\right)\right),
    \label{eq:eom_sum_over_n}
\end{equation}
or over $m$
\begin{equation}
    \dot{\overline{z}}_j = \frac{1}{2L} \sum_{k} \Gamma_k \sum_{n=-\infty}^{\infty} \cot\left(\pi i\left(\frac{z_j - z_k}{L} - n\right)\right).
    \label{eq:eom_sum_over_m}
\end{equation}
Equations (\ref{eq:eom_sum_over_n}) and (\ref{eq:eom_sum_over_m}) correspond physically to an infinite number of strips, periodic in the horizontal and vertical directions, respectively \citep{Aref2003}. 
We can then use the trigonometric identity
\REV{\begin{equation}
    \cot(a + ib) = \frac{\sin(2a) - i\sinh(2b)}{\cosh(2b) - \cos(2a)}
    \label{eq:cot_identity}
\end{equation}}
to extract $\dot{x}_j$ from equation (\ref{eq:eom_sum_over_m}) and $\dot{y}_j$ from equation (\ref{eq:eom_sum_over_n}). 
This results in the final equations of motion for the doubly periodic domain
\begin{equation}
    \begin{pmatrix}
        \dot{x}_j \\ \dot{y}_j
    \end{pmatrix} = \frac{1}{2L} \sum_{k=1}^{N_v} \phantom{}^{'} \Gamma_k \sum_{n=-\infty}^{\infty} \begin{pmatrix}
        -S_n\left(\frac{2\pi}{L}(y_j - y_k), \frac{2\pi}{L}(x_j - x_k)\right) \\ S_n\left(\frac{2\pi}{L}(x_j - x_k), \frac{2\pi}{L}(y_j - y_k)\right)
    \end{pmatrix},
    \label{eq:eom_final}
\end{equation}
where
\REV{\begin{equation}
    S_n(a,b) = \frac{\sin( a )}{\cosh(b - 2\pi n) - \cos(a)}.
\end{equation}}
The reason for extracting $\dot{x}_j$ from equation (\ref{eq:eom_sum_over_m}) rather than equation (\ref{eq:eom_sum_over_n}) is due to the need to numerically truncate the summations.
Consider for example the equilibrium configuration of $N_v = 4$ point vortices with equal circulation magnitude, equispaced in the periodic domain.
Truncating at finite $n$ results in two unmatched ghost vortices in the final vertically periodic strip in the summation, which causes a very small, but non-zero induced velocity in both the $x$ and $y$ directions.
However, this induced numerical `drift' in $x$ is substantially weaker than in $y$.
The same argument can be made for extracting $\dot{y}_j$ from equation (\ref{eq:eom_sum_over_n}), where the number of horizontal strips is truncated.

These equations of motion with $L = 2\pi$ match those derived via the method of Laplace transforms in \citet{weiss1991} (up to a rescaling of time to match their conventions). 
\citet{weiss1991} also derived the Hamiltonian for the doubly periodic system of point vortices
\begin{equation}
    H = -\frac{1}{4\pi} \sum_{j,k = 1}^{N_v} \phantom{}^{'} \frac{\Gamma_j\Gamma_k}{2} h\left( \frac{2\pi}{L}(x_j - x_k), \frac{2\pi}{L}(y_j - y_k)  \right),
    \label{eq:h_doublyperiodic}
\end{equation}
where
\REV{\begin{equation}
    h(a,b) = \sum_{m=-\infty}^{\infty} \log{\left( \frac{ \cosh(a - 2\pi m) - \cos(b) }{ \cosh(2\pi m) } \right)} - \frac{a^2}{2\pi}.
    \label{eq:h_dp_interaction}
\end{equation}}
This Hamiltonian is clearly even in both $x$ and $y$. It is clearly periodic in $y$
-- periodicity in $x$ can be shown by manipulation of the sum truncated at large $M$, before taking  the $M\to \infty$ limit.
The periodic boundary conditions break the rotational symmetry of the unbounded domain, leaving two additional constants of motion,
\begin{align}
        P_x = \sum_{j=1}^{N_v} \Gamma_j x_j, \quad P_y = \sum_{j=1}^{N_v} \Gamma_j y_j,
    \label{eq:momenta}
\end{align}
arising from the translational symmetry in $x$ and $y$, respectively. 

We solve equation (\ref{eq:eom_final}) using a differentiable point vortex solver \texttt{JAX-PV}, developed in our previous work \citep{Cleary2023}. 
This numerical solver is built on the \texttt{JAX} library \citep{jax2018github},
to allow for efficient computation of the gradient of the time-forward map $\mathbf{f}^t(\mathbf{x})$ of equation (\ref{eq:eom_final}), where $\mathbf x := (x_1, y_1, \dots, x_{N_v}, y_{N_v})$, with respect to $\mathbf x$.
The \texttt{JAX} framework also brings efficiency benefits such as just-in-time compilation and auto-vectorisation.
Time integration is performed with the symplectic second order Runge-Kutta scheme at a fixed time step of $\delta t = 10^{-3}$.

\section{Continuation of Kolmogorov RPOs}
\label{sec:forced_turb}

\subsection{Continuation in \Rey}
\label{sec:continuation}

In this section, an arclength continuation effort of the library of RPOs in two-dimensional, doubly periodic Kolmogorov flow is presented. 
As discussed in \S\ref{sec:comp_kf}, the starting library of RPOs for the continuation comprises of 174 solutions at $\Rey = 40$ and 151 solutions at $\Rey = 100$. 
Each of these solutions are defined by the state vector $\bm{X} = [\omega, s, T]^T$, where $\omega$ is a vorticity snapshot along the RPO, $s$ is the required translational shift and $T$ is the period of the RPO.
To begin the continuation of a branch, a starting solution $\bm{X}(\Rey_0)$ at $\Rey_0$ is naively perturbed in $\Rey$ by $\delta \Rey$, $\Rey_1 = \Rey_0 + \delta \Rey$. 
Using $\bm{X}(\Rey_0)$ as the initialisation at $\Rey_1$, this solution is converged via a standard Newton-GMRES-Hookstep solver to yield $\bm{X}(\Rey_1)$.
This is robust as long as $\delta \Rey$ is small, but cannot work when there are turning points in the solution branch.
For this reason, the continuation proceeds via arclength continuation after the first successful convergence.
The branch is then instead parameterised by its arclength $r$, which increases monotonically along the branch. 
The state vector is extended to include $\Rey$, i.e. $\bm{X}(r) = [\omega, s, T, \Rey]^T$, along with an additional constraint
\begin{equation}
    \frac{\partial \bm{X}}{\partial r} \cdot \frac{\partial \bm{X}}{\partial r} = 1,
    \label{eqn:re_constraint}
\end{equation}
to match the additional unknown.
The next step size in arclength $\delta r$ is controlled by
\begin{equation}
    \delta r_{i-1} = \sqrt{(\bm{X}(r_{i-1}) - \bm{X}(r_{i-2}))^2},
    \label{eq:arc_length_step}
\end{equation}
so that $\bm{X}(r_i = r_{i-1} + \delta r_{i-1})$ is computed via the modified Newton-GMRES-Hookstep solver with the extra constraint (\ref{eqn:re_constraint}). 
The initialisation for $\bm{X}(r_i)$ was set by linearly extrapolating along $r$ from the previous two states along the branch $\bm{X}(r_{i-2})$ and $\bm{X}(r_{i-i})$.
The methodology matches that presented in \citet{Chandler2013}, and full details can be found in the appendix of that paper.

The continuation algorithm was automated in the following ways. 
For each branch, the initial perturbation in $\Rey$ was set to $\delta Re = 2.5$. 
If the perturbed solution was not converged by the standard Newton-GMRES-Hookstep solver, $\delta \Rey$ was halved repeatedly until a convergence was found, up to a maximum of 4 halvings.
After this first successful step in $\Rey$, the step size in arclength $\delta r$ was set according to equation (\ref{eq:arc_length_step}), and $\bm{X}(r_i)$ was initialised using the two existing states along the branch. 
If the initialised state was not converged by the modified Newton-GMRES-Hookstep solver, then $\delta r$ was halved repeatedly until a convergence was found, again up to a maximum of 4 halvings.
The number of halvings of $\delta r$ required for convergence were recorded along the branch, as a proxy for the ease of convergence along the branch. 
If both previous solutions along the branch did not require any halvings then $\delta r$ was doubled, to speed up the continuation.
Convergence of a solution was deemed to have failed if either (1) more than 100 Newton iterations were required (2) more than 10 Hooksteps were required more than 3 times. 
Continuation of a branch was also terminated if either (1) more than 50 solutions were converged along that branch, \JP{a restriction added due to many curves repeatedly folding, or} (2) $\delta r$ was halved more than 4 times for the current solution. 
The starting $\Rey = 40$ RPOs were continued upwards in $\Rey$, while the starting $\Rey = 100$ were continued in separate computations both upwards and downwards in $\Rey$.


Long trajectories of Kolmogorov flow were also simulated at various $\Rey \in [30, 1000]$ to compute reference turbulence statistics. 
We use the probability density function (PDF) of the dissipation in these long computations as a simple measure of the dynamical significance of the UPOs to the turbulence, labelling a solution as `dynamically relevant' if its mean dissipation rate $D$ lies between the $1^{st}$ and $99^{th}$ \REV{percentile} of the reference PDF.
This method is chosen for its simplicity and ease of evaluation, but will lead to some mislabelling. Better observables should be considered in future work \citep[see e.g. the strategies in][]{Krygier2021,page2022recurrent}.
Each long trajectory was simulated for $10^5$ advective time units, and sampled at intervals of 1 advective time unit. 
Contour plots of the probability density functions (PDFs) of the dissipation of these trajectories are shown in various figures throughout the paper to aid identification of the turbulent attractor,  with boundaries of the dynamically important region of the turbulent attractor indicated by dashed black lines.

\begin{figure}
  \centerline{\includegraphics[width=\linewidth]{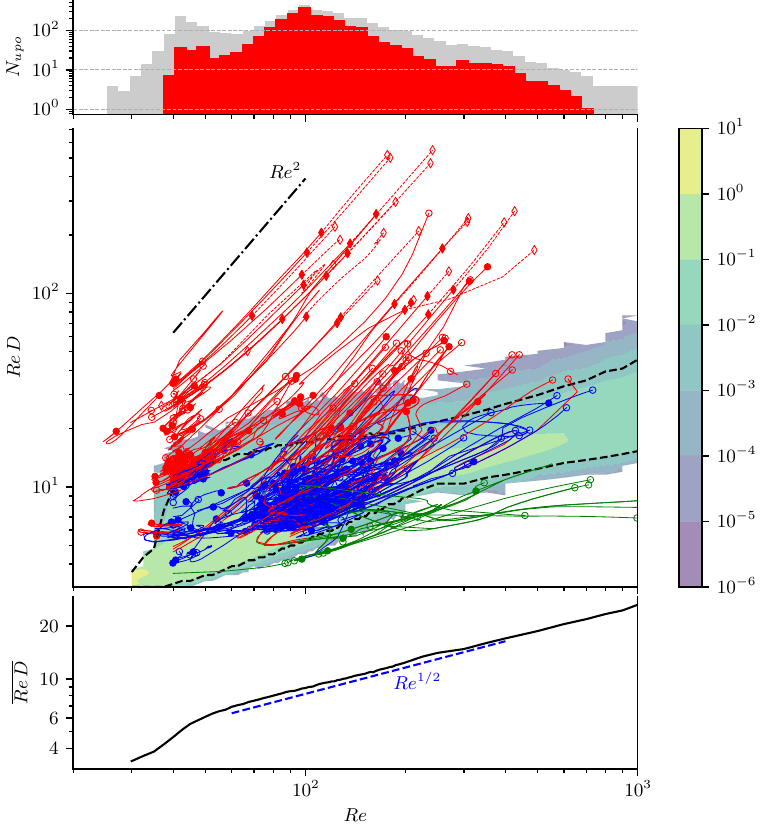}}
  \caption{
  Continuation of periodic orbits and their overlap with the turbulent dissipation PDF.
  (Top) Histograms of the number of monotonic-in-$\Rey$ subsections $N_{upo}$ of all solution branches (grey) and monotonic-in-$\Rey$ subsections fully within the dynamically important region (red) as function of $\Rey$. Both histograms are computed using 50 bins, spaced logarithmically over the range $\Rey \in [20, 1000]$.
  (Middle) The time-averaged dissipation rate $\Rey \, D$ against $\Rey$ of the arclength continuation of the initial library of UPOs starting at both $\Rey = 40$ and $\Rey = 100$. 
  \REV{The contour plot shows the reference PDFs of dissipation rate, with the $1^{st}$ and $99^{th}$ percentiles indicated by the dashed black lines.}
  The red/blue/green branches indicate those with terminal UPOs above/within/below this dynamically important region. Circles denote the terminal solution along each branch. Filled circles denote the branches which were terminated as 50 states were converged, while empty circles denote the branches which could not be continued further. 
  (Bottom) The time averaged, scaled dissipation rate $\overline{ \Rey \, D }$ of a long-time simulation is shown as a function of $\Rey$ (black), as well as the scaling law $\Rey^{1/2}$ (blue).
  }
\label{fig:classes_upo}
\end{figure}

The time-averaged dissipation rate for the RPOs along each branch was computed and is plotted in figure \ref{fig:classes_upo} over the background turbulent PDFs. 
This quantity was chosen to reveal the scaling of the volume-averaged velocity gradients with $\Rey$ for each branch.
The scaling of the time-and-volume-averaged dissipation, scaled by $Re$, $\overline{\Rey \, D}$, of the fully turbulent flow is also shown in the bottom panel of figure \ref{fig:classes_upo} and matches the `asymptotic' scaling of $\Rey^{1/2}$ identified by \citet{Chandler2013} (their observations held for $Re\leq 200$). 
The majority of branches seeded from RPOs at $\Rey = 100$ do not leave the region $\Rey \in [60, 200]$, with the branches repeatedly turning (this was the reason for terminating many continuations at 50 steps).

In the top panel of figure \ref{fig:classes_upo} the number of monotonic-in-$\Rey$ subsections of solution branches is tracked -- this is a proxy for number of unique RPOs in a particular $\Rey$ interval. The number of dynamically important solutions decreases more sharply with $\Rey$ than the total number of monotonic subsections.
\JP{
This departure of the RPOs from `dynamically relevant' dissipation values could be attributed to a number of causes. 
One is the (unlikely) exit from the attractor of RPOs via boundary crises. 
Alternatively, the RPOs may remain in the attractor but with an increasingly rare probability of visits from a turbulent orbit, hence it is deemed dynamically irrelevant under our scalar-observable based criterion. 
Finally, there is the possibility that the solutions were not in the attractor when they converged, but close by. 
This is not something that is obvious from low-dimensional projections which are typically used to assess the closeness of an RPO to the turbulence, but does seem to be borne out by the analysis below. 
}

To facilitate a discussion about the $\Rey \to \infty$ behaviour of the solutions, the branches are are labelled as belonging to one of three classes according to the dynamical importance of the final RPO on each branch, as assessed by comparison to the turbulent PDF described above. 
The final RPO of branches in class (1) have a larger dissipation rate than the dynamically important region, class (2) solutions lie in the dynamically important region (many remain confined to a small range of $Re$) and class (3) solutions have a smaller dissipation rate than the dynamically important region. 
These three classes are coloured in red, blue and green, respectively in figure \ref{fig:classes_upo}. 
At high $\Rey$, the average dissipation rate of RPOs in class (2) scales similarly to the turbulent attractor, while those in class (1) and class (3) scale more strongly (most with a scaling $D\sim Re$) or weakly ($D \sim 1/Re$ -- though we are lacking large numbers of solutions in this regime), respectively.
Note that the vast majority of solutions in classes (1) and (3) appear to continue monotonically with $Re$ after they leave the dynamically relevant region.  

\begin{figure}
  \centerline{\includegraphics[width=\linewidth]{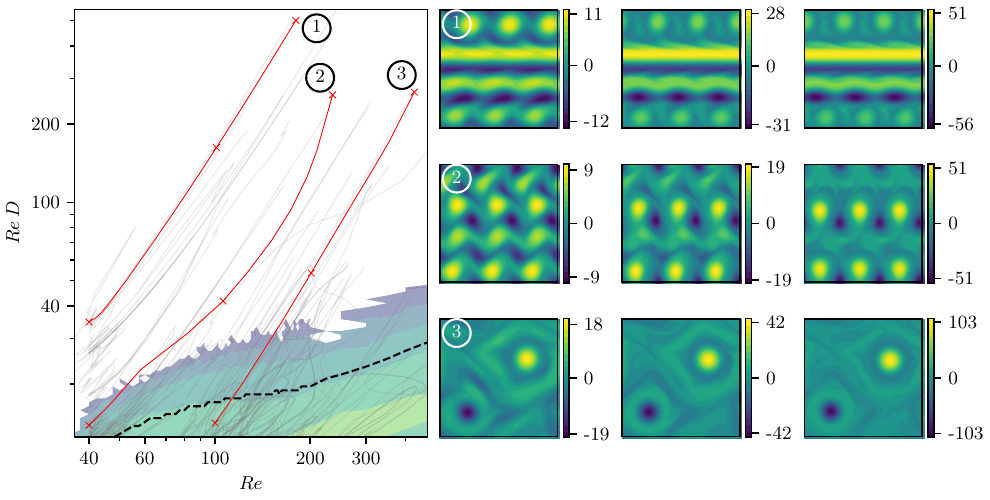}}
  \caption{The time-averaged, scaled dissipation rate $\Rey \, D$ against $\Rey$ for three representative branches in class (1). Three UPOs along each branch are sampled (indicated by crosses) and visualised on the right of the figure. Vorticity field snapshots for each sampled UPO along each branch are shown in a row, increasing in arclength from left to right. Colour maps are held constant per row to highlight the development of the vorticity gradients with $\Rey$. Note the different colour bars for each subfigures, highlighting the strengthening vortical structures.}
\label{fig:class1_eg}
\end{figure}

\JP{\subsection{Class 1: Connections to exact unforced Euler solutions}}
\JP{Some example branches from solution class 1 are shown in more detail in figure \ref{fig:class1_eg}.}
The RPOs in this class have stronger dissipation rates than the turbulence, or equivalently larger enstrophy $\langle \omega^2 \rangle_V$.
As a result, class (1) solutions have very defined vortical structures -- e.g. see the horizontal bars and clear vortex cores in figure \ref{fig:class1_eg}. 
The strength of the local vorticity associated with these features increases dramatically as $\Rey$ is raised.
For example, the maximum vorticity nearly triples for the last two points highlighted in the second row of figure \ref{fig:class1_eg}, while $\Rey$ increases from approximately 100 to 220.

\begin{figure}
    \centering
    \includegraphics[width=\linewidth]{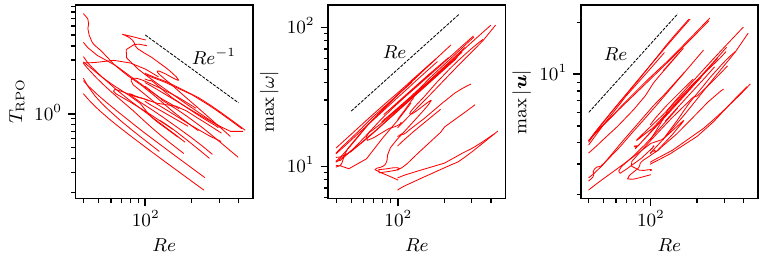}
    \caption{\REV{The scalings with $Re$ of the periods $T_{\text{RPO}}$ (left), maximum vorticity values $\max |\omega|$ (middle) and maximum velocity values $\max |\bm u|$ (right) for a subset of branches in class 1.} 
    }
    \label{fig:class1_scalings}
\end{figure}

\JP{The increasingly strong vorticity fields in class 1 meant that the baseline resolution was often inadequate as the continuation proceeded upwards in $Re$.}
\JP{As a result, some of the initial computations terminated due to a failure to converge before reaching 50 states.}
\JP{These branches are identified with filled diamonds in figure \ref{fig:class1_eg}, and were continued further at this point by doubling the number of Fourier modes in both spatial directions.
For solutions which began at $Re=40$, it was sometimes necessary to double the resolution again (to a maximum of $512\times 512$) to follow the branch further upwards in $Re$.
}

\JP{
The dissipation scaling $D \sim Re$ is consistent with an enstrophy $\langle \omega^2 \rangle_V \sim Re^2$, which implies $\omega = O(Re)$ if the vortex cores do not shrink. 
This scaling is confirmed for a number of branches in class 1 in figure \ref{fig:class1_scalings}, and the velocity field (largely induced by the high amplitude vortex cores) also follows this scaling. 
In addition, the $Re$-dependence of the period of the RPOs is also shown in figure \ref{fig:class1_scalings}, and scales like $T \propto 1/Re$. 
These scalings motivate the introduction of new variables $\Omega := \omega /Re$, $\mathbf U := \bm u /Re$ and $\tau := Re \, t$, from which equation (\ref{eq:kf_eq}) becomes
\begin{equation}
    \partial_{\tau}\Omega + \mathbf U \cdot \boldsymbol{\nabla}\Omega = \frac{1}{Re^2}\left(\Delta \Omega - n\cos ny\right).
    \label{eqn:scaled_vort}
\end{equation}
An asymptotic solution would then be sought in the form of a regular perturbation series:
\begin{align}
    \Omega(\bm x, \tau) &= \Omega_0(\bm x, \tau) + \frac{1}{Re^2}\Omega_1(\bm x, \tau) + \cdots, \\
    \mathbf U(\bm x, \tau) &= \mathbf U_0(\bm x, \tau) + \frac{1}{Re^2}\mathbf U_1(\bm x, \tau) + \cdots.
\end{align}
At each order $\Delta \Psi_i = -\Omega_i$, where $\Psi_i$ is the streamfunction associated with $\mathbf U_i$.
The dynamics at leading order are therefore governed by the inviscid vorticity equation (the curl of the Euler equation)
\begin{equation}
    \partial_{\tau} \Omega_0 + \mathbf U_0\cdot \boldsymbol{\nabla}\Omega_0 = 0,
\end{equation}
while the first order correction is
\begin{equation}
    \partial_{\tau} \Omega_1 + \mathbf U_0\cdot \boldsymbol{\nabla}\Omega_1 + \mathbf U_1\cdot \boldsymbol{\nabla}\Omega_0 = \Delta \Omega_0 - n\cos ny. 
\end{equation}
The existence of a solution to the first order problem requires that a series of solvability conditions are satisfied, 
\begin{equation}
    \int_0^T \langle \zeta_j \left(\Delta \Omega_0 - n\cos ny \right)\rangle_V d\tau = 0,
\end{equation}
where -- as shown by \citet{Zhigunov2023} -- the $\{\zeta_j\}$ are eigenfunctions associated with continuous symmetries of the $T-$(relative) periodic Euler solution $\Omega_0(\mathbf x, \tau)$. 
This is a generalisation of the solvability condition in \citet{Okamoto1994} for steady solutions, which is identical to the physical constraint derived by \citet{Batchelor1956}.
}


\JP{
The scalings of class 1 solutions confirmed in figure \ref{fig:class1_scalings} confirm this asymptotic structure and indicate that many Kolmogorov flow solutions found as low as $Re=40$ -- which have a structure at low-$Re$ that depends intimately on the forcing -- connect directly to solutions of the unforced Euler equation. 
The wide range of flow structures found here, some of which were highlighted in figure \ref{fig:class1_eg}, indicate that the connection is possible for a wide variety of flow states beyond the `unimodal' states discussed in previous work \citep{Kim2015,Kim2017,Zhigunov2023}.
The scaling of this solution class relative to the turbulence would seem to suggest that these states, while `dynamically relevant' by our definition at the $Re$ values where they were converged and potentially similar to turbulent trajectories, were outside of the turbulent attractor.
}

\JP{\subsection{Classes 2 and 3}}
\begin{figure}
  \centerline{\includegraphics[width=\linewidth]{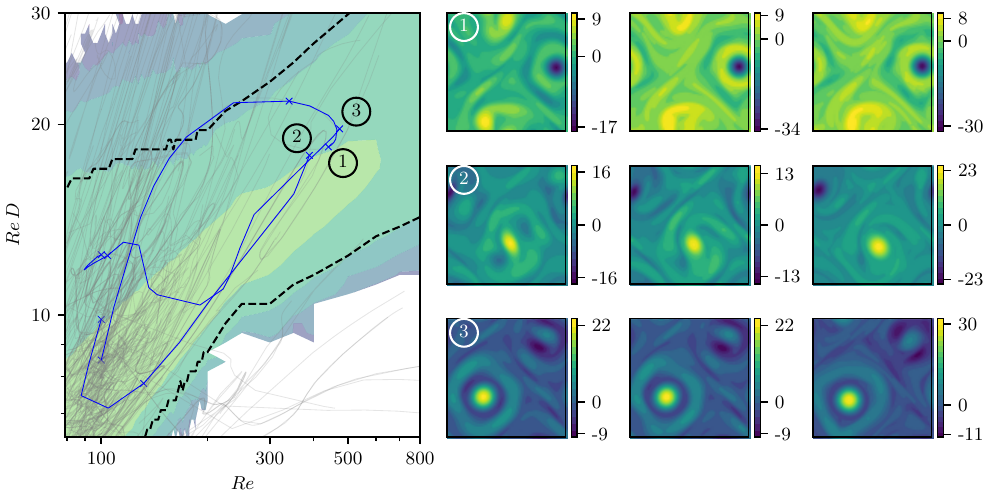}}
  \caption{The time-averaged, scaled dissipation rate $\Rey \, D$ against $\Rey$ for three representative branches in class (2). Three RPOs along each branch are sampled (indicated by crosses) and visualised on the right of the figure. Vorticity field snapshots for each sampled RPO along each branch are shown in a row, increasing in arclength from left to right.}
\label{fig:class2_eg}
\end{figure}

\begin{figure}
  \centerline{\includegraphics[width=\linewidth]{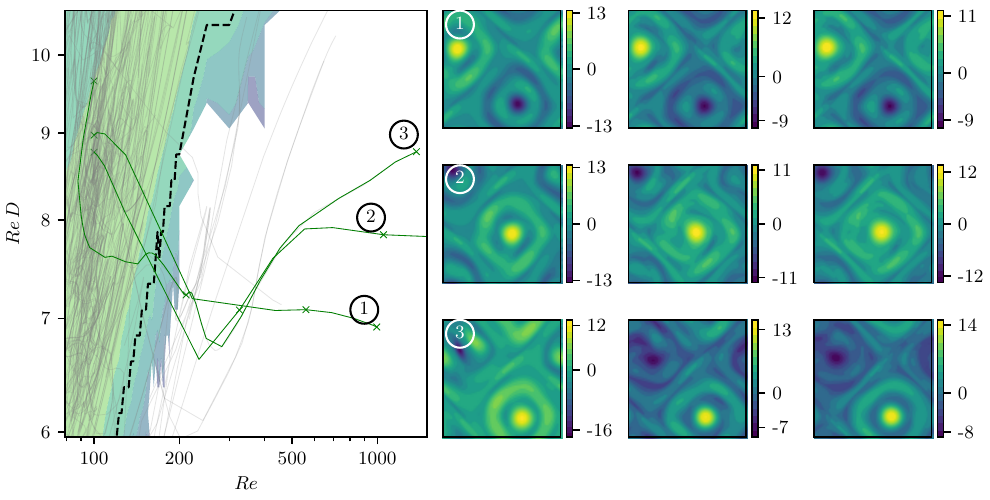}}
  \caption{The time-averaged, scaled dissipation rate $\Rey \, D$ against $\Rey$ for three representative branches in class (3). Three RPOs along each branch are sampled (indicated by crosses) and visualised on the right of the figure. Vorticity field snapshots for each sampled RPO along each branch are shown in a row, increasing in arclength from left to right. }
\label{fig:class3_eg}
\end{figure}
The RPOs in classes (2) and (3) tend to be dominated by two oppositely signed vortices, arranged similarly to the vortex condensates typically seen in high $\Rey$ and Euler solutions \citep{Zhigunov2023,Page2024}. 
The large-scale opposite-signed vortex pair in class (3) branches is less distinct from the background vorticity than in class (2) branches -- class (3) solutions still appear to show evidence of the $n=4$ forcing wave.
Multipolar structures \citep{carton1989,Morel1994} are observed in in the class (2) solutions. For example, note the co-rotating same-signed vortex `bound state' in the top row of snapshots in figure \ref{fig:class2_eg}, and the `tripole' structures (central core with two opposite-signed satellites) in the second and third rows of figure \ref{fig:class2_eg}.
In the middle row (solution `2') in figure \ref{fig:class2_eg} the tripole structure weakens with increasing $\Rey$, and the vorticity field becomes dominated by the opposite-signed large scale vortices.
In contrast to the class (2) solutions, the vorticity strength of solutions in class (3) remains practically constant, even as $\Rey$ is increased by an order of magnitude from $\Rey = 100$ to $\Rey \approx 1000$.
The dominant structures visible in the solutions in class (3) also vary remarkably little with these large changes in $\Rey$. 

\begin{figure}
    \centering
    \includegraphics[width=\linewidth]{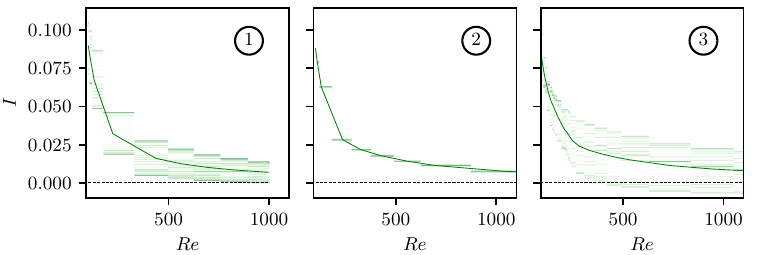}
    \caption{\REV{The full range of production $I$ at each $Re$ for the three furthest continued class (3) solutions. A logarithmic colourmap is used, and the numbering corresponds to the numbered branches in figure \ref{fig:class3_eg}.}}
    \label{fig:class3_prod}
\end{figure}

\JP{
The apparent asymptotic behaviour of the class (3) solutions, with $D\sim 1/Re$, is consistent with vorticity fields $\omega = O(1)$ (as shown in figure \ref{fig:class3_eg}). 
We also observe that $T_{\text{RPO}} \approx \text{constant}$ as $Re$ increases (not shown). 
This behaviour is consistent with a connection to solutions of a \emph{forced} Euler equation.
To see this, consider a regular perturbation expansion $(\omega, \bm u) = (\omega_0, \bm u_0) + Re^{-1}(\omega_1, \bm u_1) + \cdots$, from which the leading order contribution satisfies
\begin{equation}
    \partial_t \omega_0 + \bm u_0 \cdot \boldsymbol{\nabla} \omega_0 = -n\cos ny.
\end{equation}
There are no dissipation mechanisms, hence for $T$-periodic solutions the forcing term must extract the same amount of kinetic energy it injects over a cycle, 
\begin{equation}
    \int_0^T I\, dt = \int_0^T\langle u_0 \sin ny\rangle_V\, dt = 0.
\end{equation}
}

\JP{
The production over a complete cycle of three class (3) RPOs is reported in figure \ref{fig:class3_prod} along the solution branches. 
For two of the branches (labelled `1' and `3') negative values of the production are realised at higher $Re$, which is consistent with a forced-Euler connection in the limit $Re\to \infty$.
Further continuation would be necessary to verify this trend.}

\subsection{New RPOs at $\Rey = 100$}
\label{sec:new_upos}

Due to the non-monotonicity of the solution branches, the original library at $Re= 100$ can be supplemented with new RPOs by slicing branches anywhere they cross $\Rey = 100$ in the arclength continuation.
As a result, 101 new RPOs at $\Rey = 100$ were converged.
The initialisations in the Newton convergence for these new RPOs were generated by linearly interpolating between the two RPOs which bracket the point where the branch crosses $\Rey = 100$. 
Some branches crossed $\Rey = 100$ more than once to yield more than one new unique RPO. 

\begin{figure}
    \centering
    \includegraphics[width=\linewidth]{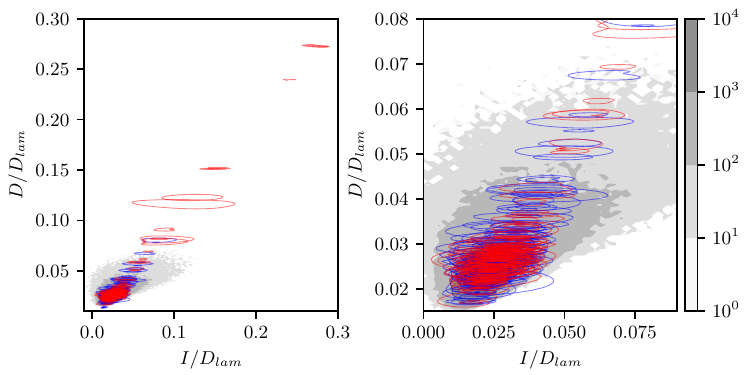}
    \caption{Energy dissipation rate $D$ against energy production rate $I$ at $\Rey = 100$, both normalised by the laminar dissipation $D_{lam} = \Rey / (2n^2)$. The 101 new RPOs at $\Rey = 100$ from the continuation are shown in red, the starting library of 151 RPOs are shown in blue. The grey background is the PDF computed from a trajectory of $10^5$ samples, separated by 1 advective time unit. The contour levels of the PDF are spaced logarithmically. The left panel shows the new very high dissipation RPOs, far from the turbulent attractor, which resulted from the continuation of the very high dissipation RPOs at $\Rey = 40$. The right panel zooms in on the turbulent attractor.}
    \label{fig:Re100_new_from_continuation}
\end{figure}

These new RPOs are reported in figure \ref{fig:Re100_new_from_continuation} in terms of their production, $I$, and dissipation $D$, normalised by the laminar value. 
Due to the localised nature in phase space of these RPOs, full coverage of the turbulent PDF at this $\Rey$ is difficult, and requires a large number of solutions.
These new RPOs go some way to increasing the coverage of the turbulent PDF, though gaps remain at higher $D$ and the tails of the $I$ marginal PDF are not covered.

\subsection{Reconstruction of turbulent statistics}
\label{sec:stats}

A variety of statistics, including full dissipation and energy PDFs, were recently successfully reconstructed via an expansion in the statistics of individual RPOs at $\Rey = 40$ by \citet{page2022recurrent}. 
In that paper, the weights in the expansion were determined from the invariant measure of a Markov chain, where the `states' on the chain were the RPOs, and the current state on a turbulent trajectory was assigned by measuring the distance to each RPO in the latent space of an autoencoder and identifying the closest solution.
We adopt a similar approach here across $\Rey \in [40, 300]$, albeit without building Markov chain models. 
Instead, RPO weights are determined by fitting to a \emph{single} representative statistic, and their robustness is then assessed by reconstructing other statistics using the same weights \citep[see the approach in][]{Redfern2024}.
We generate the various libraries of RPOs at each target $\Rey$ by slicing the solution branches from the arclength continuation, as described in section \ref{sec:new_upos}.
This allows us to track how the weights of each RPO change as $\Rey$ is increased and assess the utility of RPOs converged at one $\Rey$ value in making statistical predictions at higher values. 

\begin{figure}
    \centering
    \includegraphics[width=0.4\linewidth]{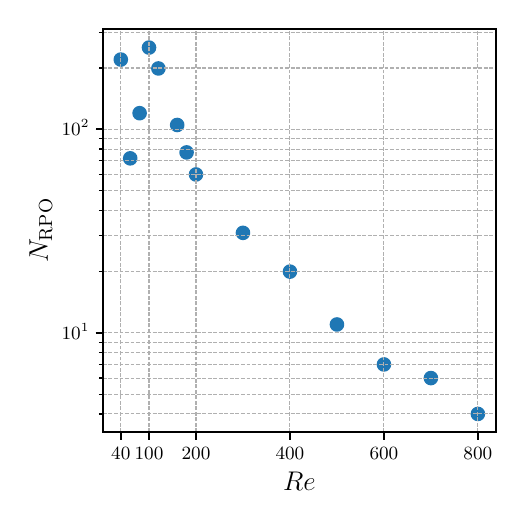}
    \caption{Number of unique RPOs converged by slicing the arclength continuation at discrete values of $\Rey$.}
    \label{fig:n_upos_per_Re}
\end{figure}

The number of unique solutions converged at each $\Rey$ considered is shown in figure \ref{fig:n_upos_per_Re}.
There is a reduction in the number of solutions reported here compared to the number of monotonic-in-$\Rey$ subsections of the solution curves reported in figure \ref{fig:classes_upo} due to multiple interpolated initial guesses in Newton sometimes converging to only one of the possible solutions on a branch \JP{repeatedly folding} back and forth (e.g. both the lower and upper branch pair at a saddle node converging to the lower-branch solution). This could be improved by reducing the arclength step sizes in the original branch continuation.
There are two distinct peaks in figure \ref{fig:n_upos_per_Re} at $\Rey = 40$ and $\Rey = 100$, as the arclength continuation was seeded from solutions at these values.
The substantial drop in numbers of solutions at the higher $\Rey$ regimes commented on in section \ref{sec:continuation} is also apparent -- there is a near-exponential drop off in the solutions.
Many of the solutions counted in figure \ref{fig:n_upos_per_Re} are dynamically unimportant and will not contribute to the reconstruction of the turbulent statistics.

We compute a fixed set of weights at a given $\Rey$, $\{ w_j(\Rey) \}_{j=1}^{N_p(Re)}$, where $N_p(Re)$ is the total number of RPOs found at that $\Rey$ value.
The ansatz, inspired by periodic orbit theory \citep{Artuso1990a,cvitanovic2013}, is that the normalised distribution $p_{\bm{w}}$ of any observable $\gamma : \mathcal{M} \to \mathbb{R}^n$ (here $\mathcal M$ is the state space) can be constructed via a linear superposition of the RPO statistics for that same observable:
\begin{equation}
   p_{\bm{w}} = \frac{\sum_{j=1}^{N_p} w_j p_{j}}{ \sum_{j=1}^{N_p} w_j },
\end{equation}
where $p_{j}(\gamma)$ is the distribution of the $j^{th}$ RPO and we are suppressing all dependencies on $\Rey$ for clarity.
We find the weights by minimising the Kullback-Leibler (KL) divergence between the reconstructed observable distribution and the distribution of the target turbulent statistic $q$
\begin{equation}
    D_{KL}(p_{\bm{w}} \| q) = \int p_{\bm{w}}(\gamma) \log \left( \frac{p_{\bm{w}}(\gamma)}{q(\gamma)}\right) d\gamma.
\end{equation}
Optimisation of the weights proceeds via gradient-based optimisation of $D_{KL}$, using the Adam optimiser \citep{Kingma2015} with an initial learning rate of $10^{-2}$. 
To ensure that the weights $w_j \in [0,\infty)$, we write them in terms of a `softplus' function, 
\begin{equation}
    w_j = \sigma_+(\hat{w}_j) := \log(1 + e^{\hat{w}_j}) \in \mathbb{R}_+,
\end{equation}
and optimise for the latent representation $\hat{w}_j \in \mathbb{R}$.
The softplus function is a smooth approximation to the typical rectifier or ReLU function and ensures that the weights are strictly positive.
The stopping criterion for convergence is set to $$ \frac{D_{KL}^i - D_{KL}^{i-1000}}{D_{KL}^i} < 10^{-8}, $$
where $D_{KL}^i$ denotes the KL divergence at the $i^{th}$ optimiser iterate. 
This optimisation amounts to fitting to a one dimensional curve and is almost instantaneous on modern hardware -- negligible compared to the cost of converging the underlying solutions. 

A key feature of the weights obtained via periodic orbit theory is that they are fixed by properties of the underlying exact solutions and do not vary between statistics.
Accordingly, robustness of the weights determined in the data-driven approach can be checked by computing the weights using a single `training' statistic, using these fixed weights to reconstruct other `test' statistics at the same $\Rey$, and then comparing KL divergences. 
Three observables are considered here as training statistics -- the normalised dissipation rate $D / D_{lam}$, normalised production rate $I / D_{lam}$ and the normalised energy $E / E_{lam}$. 

\begin{figure}
    \centering
    \includegraphics[width=\linewidth]{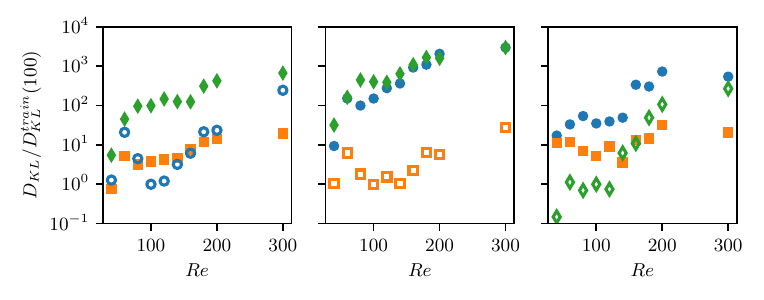}
    \caption{$D_{KL}$ loss for the reconstruction of $D / D_{lam}$ (blue circles), $I / D_{lam}$ (orange squares) and $E / E_{lam}$ (green diamonds) as a function of $\Rey$, each normalised by $D_{KL}$ at $\Rey = 100$ of the training observable, which was set to $D$, $I$ and $E$ respectively in each panel, from left to right. Unfilled markers denote this training observable in each panel. The weights were then fixed when reconstructing the distributions of the other two test observables. The normalisation constants are the laminar dissipation rate $D_{lam} = \Rey / (2n^2)$ and the laminar energy $E_{lam} = \Rey^2 / (4n^4)$.}
    \label{fig:kls}
\end{figure}
The KL divergence of the three observables are shown as a function of $\Rey$ in figure \ref{fig:kls}, setting the training observable (unfilled markers) to one of the three statistics in each panel. 
The KL divergences are normalised by the KL divergence of the training observable at $\Rey = 100$.
Both $D$ and $E$ (left and right panels in figure \ref{fig:kls}) as training statistics lead to relatively robust reconstruction of statistics in the range $40\lesssim \Rey \lesssim 150$, with a noticeable increase in $D_{KL}$ for all PDFs beyond this point.
This observation is consistent with the decreasing number of dynamically relevant solutions estimated in figure \ref{fig:classes_upo}.
The least robust observable appears to be $I / D_{lam}$, exhibiting the largest discrepancy between the training and test reconstructions of typically two orders of magnitude across all values of $\Rey$. 
The reconstruction above $\Rey = 300$ is not shown due to the lack of dynamically important solutions. 

\begin{figure}
    \centering
    \begin{minipage}{0.6\textwidth}
        \centering
        \includegraphics[width=\linewidth]{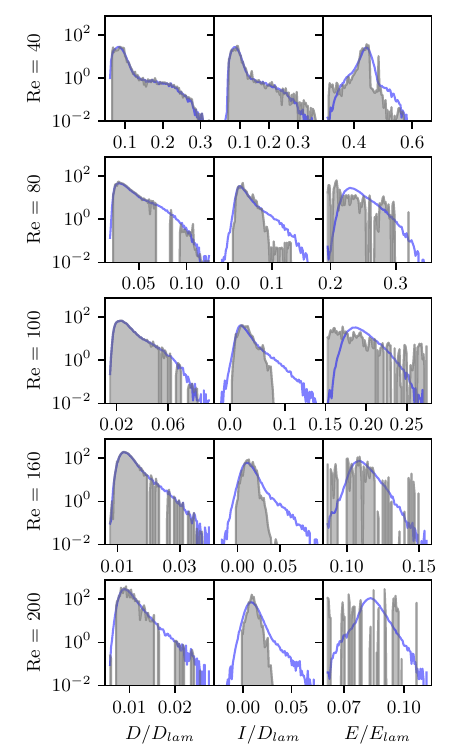}
    \end{minipage}%
    \begin{minipage}{0.4\textwidth}
        \centering
        \includegraphics[width=\linewidth]{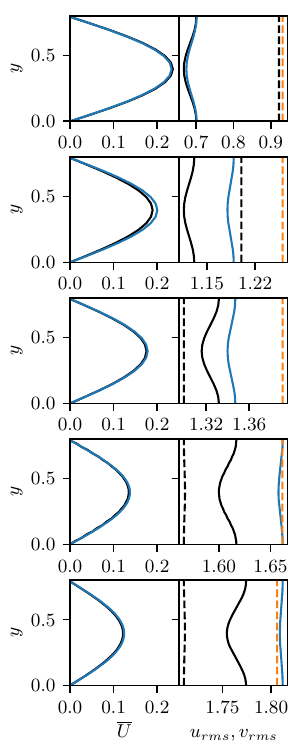}
    \end{minipage}
    \caption{Reconstruction of turbulence statistics for increasing $\Rey$ from the top row to the bottom row, with weights obtained using $D / D_{lam}$ as the training observable. (Left) From left to right, the reconstructed normalised dissipation rate $D / D_{lam}$, normalised production rate $I / D_{lam}$, and normalised energy $E / E_{lam}$ are shown in grey, while the true turbulent distributions are shown in blue. (Right) 
    In the left panels, the true turbulent mean velocity profile $\overline{U}(y)$ is shown in blue, while the reconstructed profile is shown in black. In the right panels, the true turbulent RMS velocity fluctuations $u_{rms}$ (blue continuous line), $v_{rms}$ (orange dotted line), averaged over the streamwise direction, discrete symmetries and time, are shown, while the reconstructed profiles are shown in black.}
    \label{fig:stats_reconstructed}
\end{figure}

Unlike more standard error metrics, the KL divergence \JP{(which is not a metric)} is not straightforward to interpret, and the performance of the reconstruction is assessed further in figure \ref{fig:stats_reconstructed} where the PDFs are compared to direct numerical simulation (DNS) ground truth, with $D$ as the training statistic.
As expected, the reconstruction of $D$ itself is fairly well represented by the RPOs -- particularly at $Re \in \{40, 100\}$ where the solutions were originally converged -- but with some gaps in the higher dissipation tails.
At other $\Rey$ values the reproduction of the high-dissipation tails deteriorates.
\JP{This can be attributed to the emerging dynamical irrelevance of solution class (1), which climb to increasingly high dissipation rates to connect to inviscid solutions and which may be outside of the turbulent attractor.}
Unsurprisingly, the reconstruction of \JP{the more probable production rates} is also fairly robust, though the high-$I$ tail is only really apparent from the RPOs at $Re=40$.
However, reconstruction of the kinetic energy PDF is noticeably poor compared to $D$ and $I$, particularly outside of the $Re$ where solutions were originally converged. 

\JP{
One striking aspect of the PDF reconstruction is the failure to generate the high-$I$ tails (and also the negative values of production observed at higher $Re$ values).
This is perhaps surprising given the relatively robust reconstruction of the dissipation and the fact that $\overline{D} = \overline{I}$ when averaging is over any relative periodic solution. 
However, note that the production PDF is significantly wider than that of dissipation. 
For instance, peak values of $I$ for $Re\gtrsim 100$ increase to more than twice the peak dissipation values (see PDFs in figure \ref{fig:stats_reconstructed}).
Therefore, a periodic orbit with time-averaged dissipation $\overline{D}$ which samples these extreme $I$ values will need to spend longer in the low-$I$ region. 
The absence of this behaviour in our solution library is therefore likely due to the relatively short periods of the converged cycles. 
}

\JP{One common feature for all reconstructed distributions is that} continuation of the RPOs from other $\Rey$-values provides coverage for the more probable events, but the solutions representing the rarer excursions to, for example, high dissipation or high energy regions of the state space appear to be relevant only in a small $\Rey$ range. 
This is true even in the `asymptotic' regime beyond $Re \gtrsim 100$ -- for instance see the relative lack of solutions in the tails at $Re=160$ in figure \ref{fig:stats_reconstructed} compared to those at $Re=100$. 

On the other hand, predictions for the mean velocity and root-mean-square (RMS) velocity fluctuations \citep[averaged over the streamwise direction, the 16 discrete symmetries of $n=4$ Kolmogorov flow and time, see][]{Chandler2013, Farazmand2016} are relatively robust over the range of $Re$ considered in figure \ref{fig:stats_reconstructed}. 
All three predictions at $\Rey = 40$ are very close to the true DNS profiles, and the mean velocity profile is predicted quite faithfully at all $\Rey$ considered.
\JP{Note the significant weakening of $\overline{U}$, which was commented on in \S \ref{sec:vort_id}, consistent with the increased decoupling from the background forcing (for reference, the laminar velocity profile is proportional to $Re$ in our scaling).}
\JP{On the other hand, the fluctuating velocities grow with increasing $Re$.} Predictions for the RMS velocity fluctuations suffer slightly for $\Rey > 40$, but errors are consistently below $\mathcal{O}(10\%)$.
These velocity profile reconstructions are roughly in line with the observable-agnostic predictions at $\Rey \in \{40, 100\}$ in \citet{page2022recurrent}.

\begin{figure}
    \centering
    \begin{minipage}{0.5\textwidth}
        \centering
        \includegraphics[width=\linewidth]{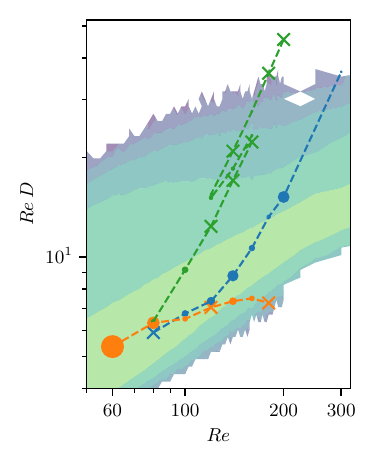}
    \end{minipage}%
    \begin{minipage}{0.5\textwidth}
        \centering
        \includegraphics[width=\linewidth]{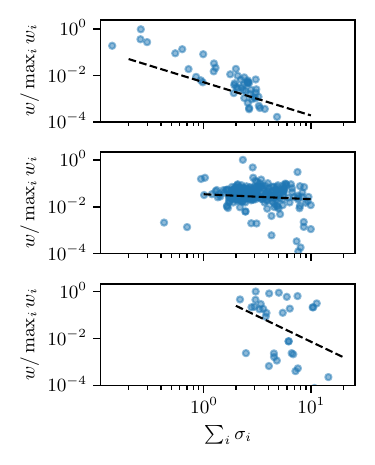}
    \end{minipage}%
    \caption{(Left) Visualisation of the changing weights along the solution curves of some example RPOs, with $D$ as the training statistic. The RPOs are visualised via their time-averaged, scaled dissipation rate, and the relative size of each circle along the curves indicates the relative weight $w_j / \max_j w_{j}$ of the RPO at that particular $\Rey$. Each colour represents a different solution branch; larger circles indicate larger weights, while crosses denote RPOs with $w_j / \max_j w_{j} < 10^{-4}$. (Right) Dependence of the weights on the real part of the sum of unstable Floquet exponents $\sum_j \sigma_j, \, \sigma_j > 0$, for each RPO at $\Rey = 40, 100, 200$ from top to bottom. The dotted black line denotes the line of best fit, indicating an inverse dependence of  $(\sum_i \sigma_i)^{-1.43}$, $(\sum_i \sigma_i)^{-0.21}$ and  $(\sum_i \sigma_i)^{-2.19}$, at $Re=40,100,200$ respectively. A total of 163, 22 and 28 RPOs are cut-off, respectively, due to the truncation of the weights at $10^{-4}$.
    }
    \label{fig:weights_diss_Re}
\end{figure}

Tracking how the weights for the RPOs change along their solution curves provides an indication of the changing dynamical importance of that solution as $Re$ increases.  
This is shown for a sample of representative solution branches in the left panel of figure \ref{fig:weights_diss_Re}, where $D$ was again used as the training statistic to fit the weights.
The relative weight $w_j / \max_j w_j$ for each RPO is indicated by the size of its marker (circles), while points where the weight $w_j / \max_j w_j < 10^{-4}$ are indicated with a cross. 
The figure indicates the same broad trends identified above, that weights are larger when the RPO sits closer to the centre of the turbulent distribution (see orange curve in figure \ref{fig:weights_diss_Re}) and drop as the RPO \JP{moves to higher dissipation values to become less dynamically relevant (see green and blue curves -- these are class (1) solutions)} -- though note that the weights tend to vary non-monotonically along the branch.

\JP{
The data-driven approach to assigning weights to solutions is driven by an acceptance that the solution library is incomplete. 
One major flaw in our incomplete library of solutions is that no search was made for pre-periodic solutions with a finite number of shift-reflects in the vertical. 
We have also seen that many solutions which were believed to be on the attractor (overlap with turbulence in low-dimensional projections) may actually be outside, but still somehow contain `turbulent’ dynamical processes.
An examination of the dependence of the weights on the Floquet multipliers of the RPOs (right panel of figure \ref{fig:weights_diss_Re}) indicates the expected inverse correlation $w_j \sim (\sum_i \sigma_i)^{-1}$ \citep[see also the discussion in][]{Redfern2024} for moderate $\Rey = 40$, but shows little correlation at higher $Re$.
The indication is a decoupling of the fitting from any underlying dynamical properties of the solutions, which is perhaps to be expected given both the fate of classes (1) and (3) as $Re \to \infty$ and the diminished cover of the tails of the PDFs. 
}

\JP{
It is pertinent at this point to question the utility of a periodic orbit statistical reconstruction in light of these results. 
The library of solutions at $Re=100$ (for example) is to our knowledge the largest set of (apparently `dynamically relevant’) solutions assembled and is the result of thousands of GPU hours of computation \citep[as documented in][]{page2022recurrent,Page2024} augmented by our continuation effort. 
The success of future efforts rests both on convergence of solutions with longer periods, bringing additional computational challenges, \emph{and} an assessment of whether these states are embedded in the turbulent attractor, where we have seen that an inspection of low-dimensional projections is likely very misleading.
}
One caveat here is that this analysis is based on two-dimensional turbulence, and may not apply directly to the three-dimensional case where the nature of the cascade is different. 
\JP{Given the apparent connection of many of the states at $Re=100$ to inviscid solutions, we explore now whether their key dynamical features can be captured in a much simpler point vortex model.}

\section{Labelling with point vortex RPOs}
\label{sec:pv_matching}
\JP{Motivated by the apparent connection of the class (1) RPOs to solutions of the unforced Euler equation, and the wide range of vortex dynamics encompassed in this set of orbits,}
this section explores the feasibility of fitting exact point vortex solutions to the RPOs in our library.
\JP{Our analysis is focused} primarily on the large set of solutions converged at $Re=100$, \JP{of which a large number achieved the class 1 $D\sim Re$ scaling on continuation, some after going through a series of fold bifurcations before moving away from `turbulent' dissipation values.
The motivation for doing this is the known connection of continuous families of Euler solution to point vortex dynamics \citep[e.g. Stuart vortices --][]{stuart1967,Crowdy2003a} in particular limits.
}

Fitting is done by matching the dynamics of the point vortices to the dynamics of the vortex cores of the turbulent RPOs, while insisting that the time evolution of the point vortices is also a relative periodic orbit.
This approach relies on gradient-based optimisation of a scalar loss function which depends on entire point vortex trajectories, and which is accomplished here using a fully differentiable solver \citep{Cleary2023}.

\begin{figure}
    \centering
    \includegraphics[width=\linewidth]{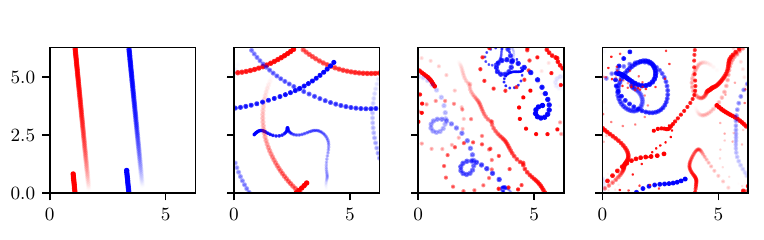}
    \caption{Sample trajectories of randomly generated point vortex systems with (from left to right) $N_v= 2,3,4$ and 6 vortices respectively in a doubly periodic domain of size $2\pi \times 2\pi$, with circulations normalised such that $\max_{\alpha} |\Gamma_{\alpha}| = 10$ and $\sum_{\alpha} \Gamma_{\alpha} = 0$. Trajectories are plotted as red (blue) points denoting positive (negative) circulation, such that increasing opacity denoting increasing time and the separation of points along a trajectory gives an indication of the speed of that vortex. Each system is simulated for 30 time units, and is set to have net zero circulation. }
    \label{fig:pv_trajs}
\end{figure}

The labelling procedure is broken down into three stages: 
(i) extraction of the vortex cores from the turbulent RPOs and the initialisation of a representative point vortex system (described in \S\ref{sec:dyn_init});
(ii) gradient-based optimisation which updates the initial positions and circulations of the point vortex system to improve the matching between the point vortex trajectories and the dynamics of the reference turbulent RPO (\S\ref{sec:convergence});
(iii) a point vortex Newton-GMRES-Hookstep RPO solver then attempts to converge to an exact point vortex RPO (also detailed in \ref{sec:convergence}).

At the initialisation stage a choice must be made for the number of point vortices $N_v$ used in the fitting of each RPO.
Representative trajectories of the doubly periodic point vortex system for various choices of $N_v$ are shown in figure \ref{fig:pv_trajs}.
The $N_v = 2$ system is integrable \REV{\citep{stremler1999}}: the vortices translate uniformly together -- see the left-most panel of figure \ref{fig:pv_trajs}.
The $N_v = 3$ system is also integrable when the net circulation of the system is zero \JP{-- the conserved quantities being the Hamiltonian and the horizontal and vertical momenta which commute for $\sum_j \Gamma_j = 0$} \REV{\citep{stremler1999}}, as is the case here to satisfy the Gauss constraint. 
This system was studied extensively by \citet{stremler1999}, who showed it can be mapped to the problem of advection of a passive particle by a system of stationary vortices, with considerably richer dynamics than the equivalent system on the unbounded plane. 
For example, they showed that if the ratio of the vortex circulations is rational, then the relative motion of the vortices is periodic, exhibiting so-called `paired', `coupled', `collective' or `wandering' motion.
In the more general case of a non-rational ratio of vortex circulations, as shown in the second panel of figure \ref{fig:pv_trajs}, the relative motion can be aperiodic. 
For higher $N_v \ge 4$, the system is not integrable and exhibits the usual features of low-dimensional Hamiltonian chaos \citep{Aref1982}.

\subsection{Vortex initialisation}
\label{sec:dyn_init}

The first stage in the labelling procedure is the initialisation of a point vortex system from the reference turbulent RPO at $\Rey = 100$. 
First, $N_{S} = 50$ snapshots are sampled from the reference Kolmogorov RPO, equally spaced in time. 
The vortex cores in each snapshot are extracted using the vortex identification method outlined in section \ref{sec:vort_id} and
the modal number of vortex cores at each snapshot throughout the period of the RPO, $\bar{N}_v$, is computed.

This procedure seeks to initialise a point vortex system which matches the strongest vortex cores over the period of the turbulent RPO. 
Trajectories of vortex cores in the reference RPO are tracked by connecting vortex cores $i$ and $j$ in sequential snapshots $k$ and $k+1$ which satisfy the similarity condition
\begin{equation}
    \frac{\| \bm{X}_i^k - \bm{X}_j^{k+1} \|_2}{\| \bm{L} \|_2} < 0.1,
\end{equation}
where $\bm{X}_i = (\Tilde{x}_i, \Tilde{y}_i, \Gamma_i, \mathcal{A}_i)$ is a state vector collating the centre of vorticity, circulation and area of each vortex core.
The normalisation $\bm{L}$ is set to $(L_x, L_y, \Gamma_i, \mathcal{A}_i)$, rather than $\bm{X}_i$.
This choice of normalisation is made so that the similarity condition is independent of the location of the vortex cores on the domain, and depends instead on the inter-core distance -- note that periodic boundary conditions are accounted for when computing the difference between $\Tilde{\bm{x}}_i^k$ and $\Tilde{\bm{x}}_j^{k+1}$.
As the turbulent solutions are typically RPOs, there is a translational discontinuity between the first and final snapshot of the period of the solution.
This discontinuity is taken into account by translating the vortex cores in the first snapshot by the reverse translational shift when attempting to connect a trajectory between the first and final snapshots. 

Some vortex cores will not satisfy the vorticity thresholding condition in equation (\ref{eq:vort_extraction_cond}) for all snapshots along their trajectories.
In general, very few vortex-core trajectories exist over the full period of the reference RPO due to the fluctuating strength of the vortices.
The trajectories of the vortex cores are sorted by their average circulation over their existence,
\begin{equation}
    \langle \Gamma_i \rangle = \frac{\int_{0}^T \Gamma_i(t) \ind_{i}(t) dt}{\int_{0}^T \ind_{i}(t) dt}, 
\end{equation}
where the indicator function $\ind_{i}(t)$ equals 1 if vortex core trajectory $i$ exists at time $t$, and 0 otherwise.
The average strength $\langle \Gamma_i \rangle$ is used as a measure of the dynamical importance of the $i^{th}$ vortex core trajectory. 
The $\bar{N}_v + 1$ most dynamically important trajectories are extracted, and $\langle A_i \rangle $ is also computed for each of these trajectories. 
The areas and circulations of the point vortex state are initialised with $\langle A_i \rangle $ and $\langle \Gamma_i \rangle $, respectively (the areas are required to compare to the reference RPO evolution -- see below).
If all of the extracted trajectories co-exist in some snapshot of the reference RPO, the centres of vorticity of the vortex cores in this snapshot are used to initialise the positions of the point vortices. 
Otherwise, the snapshot with the smallest number of absent vortices is used and the positions of the absent vortices are initialised with their average centres of vorticity $\langle \tilde{\bm{x}}_i \rangle$.

\subsection{Convergence}
\label{sec:convergence}

The dynamical initialisation algorithm outlined in the previous subsection outputs the position, circulation and area of $\bar{N}_v + 1$ point vortices.
These point vortices correspond to the $\bar{N}_v + 1$ strongest vortex core trajectories extracted from the reference RPO.
However, there is no guarantee that this initialisation will be near to an exact point vortex solution.
The initialisation is also very sensitive to the vorticity extraction condition in equation (\ref{eq:vort_extraction_cond}).
This is because vortex cores will not satisfy equation (\ref{eq:vort_extraction_cond}) for sections of their trajectories, directly impacting the modal number of vortex cores extracted $\bar{N}_v$.
For these reasons, we employ gradient-based optimisation to improve the point vortex fit and reduce the sensitivity to the extraction method, before converging onto an exact point vortex RPO with a Newton-GMRES-hookstep solver.

The gradient-based optimisation seeks to enforce two specific behaviours in the point vortex evolution:
(i) that the point vortices must shadow the vortex cores from the reference RPO closely and
(ii) that the point vortex evolution itself must also be (relative) time periodic. 
We construct a scalar loss function $\mathscr L$ to target these two effects.
Unlike in the Newton-GMRES-hookstep solver subsequently used to converge exact solutions, the circulations of the point vortices are allowed to change during optimisation. 

Matching to the reference vortex cores is done by a contribution $\mathscr L_{\text{match}}$ to the total scalar loss function $\mathscr L$.
The contribution $\mathscr L_{\text{match}}$ is included to attempt to minimise the difference between the reference RPO and a grid-based representation of the point vortex solution.
The singular point vortex circulations of an $N_v$-vortex configuration at position $\mathbf x^* = [\bm{x}^*_1, \dots, \bm{x}^*_{N_v}] \in \mathbb{R}^{2 N_v}$ (note $\bm x_{\alpha}^* := [x_{\alpha}^*, y_{\alpha}^*]$), with circulations $\bm{\Gamma} \in \mathbb R^{N_v}$ and areas $\bm{\mathcal A} \in \mathbb R^{N_v}$, are used to define a doubly periodic observable by placing a Gaussian at each vortex location:
\begin{equation}
    G(\bm{x}; \mathbf x^*, \bm \Gamma) := \sum_{\alpha=1}^{N_v} g_\alpha(\bm{x}; \bm x^*_{\alpha}, \Gamma_{\alpha}),
    \label{eq:gauss_phys}
\end{equation}
where
\begin{equation*}
    g_\alpha(\bm{x}; \bm x^*_{\alpha}, \Gamma_{\alpha}) := \sum_{n,m = -\infty}^{\infty} \Tilde{g}_{\alpha}(\bm{x}; \bm x^*_{\alpha} + (nL, mL), \Gamma_{\alpha})
\end{equation*}
is a doubly periodic function with Gaussian contributions from all image vortices, i.e.
\begin{equation*} 
    \Tilde{g}_{\alpha}(\bm{x}; \bm x^*_{\alpha}, \Gamma_{\alpha}) = \frac{\Gamma_{\alpha}}{2\pi \sigma_{\alpha}^2} \exp \left(\frac{-(\bm x - \bm x^*_{\alpha})^2 }{2\sigma_{\alpha}^2}\right).
\end{equation*}
In this expression, the properties of the vortex cores extracted from the reference turbulent RPO are used to define the standard deviation of the Gaussian, $\sigma_{\alpha}^2 = 0.1 \mathcal A_{\alpha}$. 
This smoothing of the vorticity delta functions centred on each point vortex allows for comparison between snapshots of the point vortex system and the vorticity field of the reference Kolmogorov RPO. 
It also allows for the possibility of vortex permutations if their strengths and areas are equal.
To compute the doubly periodic Gaussian distribution in (\ref{eq:gauss_phys}), we approximate its Fourier series with the Fourier transform of the Gaussian distribution on the infinite domain
\begin{equation}
     \frac{1}{4\pi^2}\int_{-\pi}^{\pi} \int_{-\pi}^{\pi} g_{\alpha}(\bm{x}) e^{-i\bm{k}\cdot \bm{x}}d^2\bm{x} \approx \frac{1}{4\pi^2}\int_{-\infty}^{\infty} \int_{-\infty}^{\infty} \Tilde{g}_{\alpha}(\bm{x}) e^{-i\bm{k}\cdot \bm{x}}d^2\bm{x} ,
\end{equation}
which we find to be numerically accurate to machine precision when $\sigma_{\alpha}^2 < 2$.
The Fourier transform approximation $\hat{g}_{\alpha}(\bm{k})$ 
is derived by applying the shift and stretching Fourier theorems to the standard Gaussian Fourier transform
\begin{equation}
    \hat{g}_{\alpha}(\bm{k}) =  \frac{1}{4\pi^2}e^{-i\bm{k}\cdot\bm{x}_{\alpha}} e^{-\bm{k}^2\sigma_{\alpha}^2 / 2} \, .
    \label{eq:gauss_fft}
\end{equation}
Taking the inverse Fourier transform of (\ref{eq:gauss_fft}) yields the approximation for the doubly periodic Gaussian distribution in (\ref{eq:gauss_phys}).

The function $\mathscr L_{\text{match}}$ then minimises the average $L_2$ distance between snapshots extracted from the turbulent RPO $\{\omega(\bm{x}, t) : 0 \leq t < T_{\text{RPO}}\}$ and the Gaussian observable of the point vortex solution over the period $T_{\text{RPO}}$ of the turbulent solution,
\begin{equation}
    \mathcal{L}_{\text{match}}(\mathbf{x}^*, \bm{\Gamma}) := \frac{1}{N_{S}} \sum_{i=1}^{N_{S}} \frac{\left\|  G(\bm{x}; \mathbf{f}^{t_i}(\mathbf{x}^*), \bm{\Gamma})  - \omega(\bm x, t_i)\right\|^2}{\left\| \omega(\bm x, t_i)\right\|^2}
    \label{eq:loss_match}
\end{equation}
where $\mathbf{f}^t(\mathbf{x}^*)$ is the time forward map of equation (\ref{eq:eom_final}), $t_n = n T_{\text{RPO}} / N_{S}$ are equally spaced times over the period and the norm $\| v \|^2 := (1/V)\int_V v^2 \, d^2 \bm x$.
The period and vortex areas $\bm{\mathcal A}$ are kept fixed throughout the optimisation, so the only adjustable parameters for $\mathscr L_{\text{match}}$ are the initial vortex positions $\mathbf{x}^*$ and their circulations $\bm \Gamma$.

The second contribution to the overall loss function $\mathscr L$ is designed to target point vortex RPOs, and is termed $\mathscr L_{\text{RPO}}$. 
To allow for permutations of the point vortices between the states $\mathbf{x}^*$ and $\mathbf{f}^{T_{\text{RPO}}}(\mathbf{x}^*)$, this loss function is also defined in terms of the periodic Gaussian observable in equation (\ref{eq:gauss_phys}),
\begin{equation}
    \mathcal{L}_{\text{RPO}}(\mathbf{x}^*, \bm{\Gamma}, \bm s) = \frac{\left\|  G(\bm{x}; \mathcal{T}^{\bm s} \mathbf{f}^{T_{\text{RPO}}}(\mathbf{x}^*), \bm{\Gamma})  - G(\bm{x}; \mathbf x^*, \bm{\Gamma})\right\|^2}{\left\| G(\bm{x}; \mathbf x^*, \bm{\Gamma})\right\|^2},
    \label{eq:loss_upo}
\end{equation}
where $\mathcal{T}^{\bm s} = \mathcal{T}^{(s_x, s_y)}$ generates translational shifts in the $x-$ and $y-$ directions. 
While the turbulent RPOs cannot shift in $y$ (we do not have solutions in our library which shift-reflect), the point vortex states are allowed to drift in both directions.
The adjustable parameters are $\mathbf{x}^*(0)$ and $\bm \Gamma$ (as in $\mathscr L_{\text{match}}$), as well as the shifts $s_x$ and $s_y$. 

The full loss function is a combination of the two terms discussed above:
\begin{equation}
    \mathscr{L}(\mathbf{x}^*, \bm{\Gamma}, \bm s) = \kappa \beta \mathscr{L}_{\text{match}}(\mathbf{x}^*, \bm{\Gamma}) + (1 - \kappa) \mathscr{L}_{\text{UPO}}(\mathbf{x}^*, \bm{\Gamma}, \bm s),
    \label{eq:total_loss}
\end{equation}
where the hyperparameter $\beta = 50$ is set to a large number to increase the relative importance of the matching loss term.
This large weighting of $\mathscr L_{\text{match}}$ places a heavy emphasis on finding point vortex solutions which closely mimic the dominant vortex dynamics, which are themselves time periodic. 
The second term $\mathscr L_{\text{RPO}}$ should be thought of as a small correction to aid this convergence towards exactly recurring solutions. 
The hyperparameter $\kappa$ is defined to conveniently set either of the loss terms to zero.

The gradients $\boldsymbol\nabla_{\mathbf{x}^*, \bm{\Gamma}, \bm s}\mathscr{L}$ are passed to an Adam optimiser \citep{Kingma2015} with an initial learning rate of $\eta = 10^{-2}$.
The optimiser is allowed to run for a maximum $N_{\text{opt}} = 1000$ steps, or until $\mathscr{L}_{\text{RPO}} \le 10^{-5}$.
The value of the hyperparameter $\kappa$ has a strong effect on the likelihood of satisfying this tolerance on $\mathscr{L}_{\text{UPO}}$. 
For example, for $\kappa = 0.5$, just 4\% of initialisations reach this tolerance \REV{(the $\mathscr{L}_{\text{match}}$ term is still heavily weighted due to our choice of $\beta$)}, compared to the 74\% of initialisations for $\kappa = 0$.
All the optimiser outputs are passed to a Newton solver to attempt convergence regardless of whether they reach this tolerance level.

Two non-differentiable operations are then performed on the output of the gradient-based optimisation.
The first is to remove any vortices which satisfy $|\Gamma_{\alpha}| < \varepsilon$, where $\varepsilon$ is a tolerance chosen as one of $\varepsilon \in \{0, 0.01, 0.05\}$.
For reference, vortex circulations were typically in the range $| \Gamma_{\alpha}| \in [1,5]$ after the gradient-based optimisation.
As $\mathcal{L}_{\text{RPO}}$ is computed in terms of the Gaussian observable, the optimiser tends to reduce the circulation of any point vortices with a large contribution to $\mathcal{L}_{\text{RPO}}$. 
Removing these weak vortices improves the final convergence rate of the Newton-GMRES-hookstep solver. 

The second non-differentiable operation is the identification of any vortices which 
\REV{permute their positions over the period of the point vortex solution}.
As $\mathcal{L}_{\text{RPO}}$ is computed in terms of the grid-based Gaussian observable, vortex cores with similar areas and circulations are allowed to permute over the period $T_{\text{RPO}}$ while maintaining small $\mathscr L_{\text{RPO}}$ loss. 
As such, we have not yet needed to be explicit about the permutation symmetry of point vortices with equal circulation.
However, the Newton-GMRES-hookstep solver used in the final convergence step finds roots of the residual defined in terms of the vortex positions
\begin{equation}
     \bm{F}(\mathbf{x}^*, \bm{s}, T ) = \mathcal{T}^{\bm{s}}\mathbb{P}\,\mathbf f^T(\mathbf{x}^*) - \mathbf{x}^*,
\end{equation}
where the permutation $\mathbb{P}$ must be computed a priori. 
A modified Jonker-Volgenant variant \citep{crouse2016} of the Hungarian algorithm \citep{kuhn1955} is used to compute the vortex permutation $\mathbb{P}$ between $\mathbf x^*$ and $\mathbf f^{T_{\text{RPO}}}(\mathbf x^*)$ after the optimisation and prior to the initialisation of the Newton solver \citep{Cleary2023}. 
The simplest permutation is a pairwise swap of two vortices $\bm{x}^*_{\alpha}$ and $\bm{x}^*_{\beta}$, which can be written as a 2-cycle, $\mathbb{P} = (\alpha \, \beta)$.
However, higher-order $k$-cycles which permute $k$ vortices are also possible in point vortex RPOs. 
In order for the state $\mathcal{T}^{\bm{s}}\mathbb{P}\,\mathbf f^T(\mathbf{x}^*)$ to be a symmetric copy of $\mathbf{x}^*$, all vortices belonging to the $k$-cycle must share the same circulation.
Any permutation can be decomposed into disjoint cycles $\xi_i$, such that $\mathbb{P} = \prod_i \xi_i$ and each vortex belongs to at most one $\xi_i$.
To ensure that the permutation symmetry is respected, the circulation of all vortices belonging to each $\xi_k$ is then manually set to their mean circulation.
All vortex circulations are then fixed and not further updated by the Newton solver.

When converging solutions with the Newton solver, we apply a threshold on the Newton residual \emph{and} a condition that the period is not significantly adjusted after optimisation:
\begin{align}
     \frac{ \left \| \mathcal{T}^{\bm{s}}\mathbb{P}\,\mathbf f^T(\mathbf{x}^*) - \mathbf{x}^*
     \right \|}{\| \mathbf{x}^* \|}  &< 10^{-8}, \label{eq:newt_resi_cond} \\
     \frac{ \left | T - T_{\text{RPO}} \right |}{ T_{\text{RPO}} }  &< 0.25.
\end{align}
The optimal initial size of the trust region for the Hookstep solver was found to be $0.01 \times \| (\mathbf x^*, \bm s, T) \|_2$.
This small trust region resulted in a greater rate of dynamically representative convergences -- larger values resulted in point vortex RPOs that bore little resemblance to the reference solution;
smaller initial trust regions required too many Newton steps for convergence. 

\subsection{Optimisation results}
\label{sec:results}
\begin{table}
  \begin{center}
    \def~{\hphantom{0}}
    \begin{tabular}{c|c|c|c|c|c|c}
        Name & $\kappa$ & $\varepsilon$ & Total Convergences & Total Matches & $\mathscr L_{\text{rel}} < 1$ \\ \hline
        1a & 1/2 & 0 &  238  &  117 &  63     \\
        1b & 1/2 & 0.01 & 237  & 117 &   63    \\
        1c & 1/2 & 0.05 & 238  &  117 &  63    \\
        2a & 0 & 0.01 &  411  &  138 &   64    \\
        2b & 0 & 0.05 & 468  & 146  &   73   \\
        3a & Anneal & 0.01 & 255  & 120  &  67    \\
        3b & Anneal & 0.05 & 266  & 122  &  68    \\\hline
    \end{tabular}
    \caption{Summary of point vortex RPO-fit experiments run using 252 turbulent RPOs at $\Rey = 100$ to initialise the optimisation. Each reference RPO yields 3 different initialisations, by extracting the modal number of vortex cores, $N_v$, as well as $N_v \pm 1$, over the period of the reference RPO.}
    \label{tab:upo_experiments}
  \end{center}
\end{table}

\begin{figure}
    \centering
    \includegraphics[width=\linewidth]{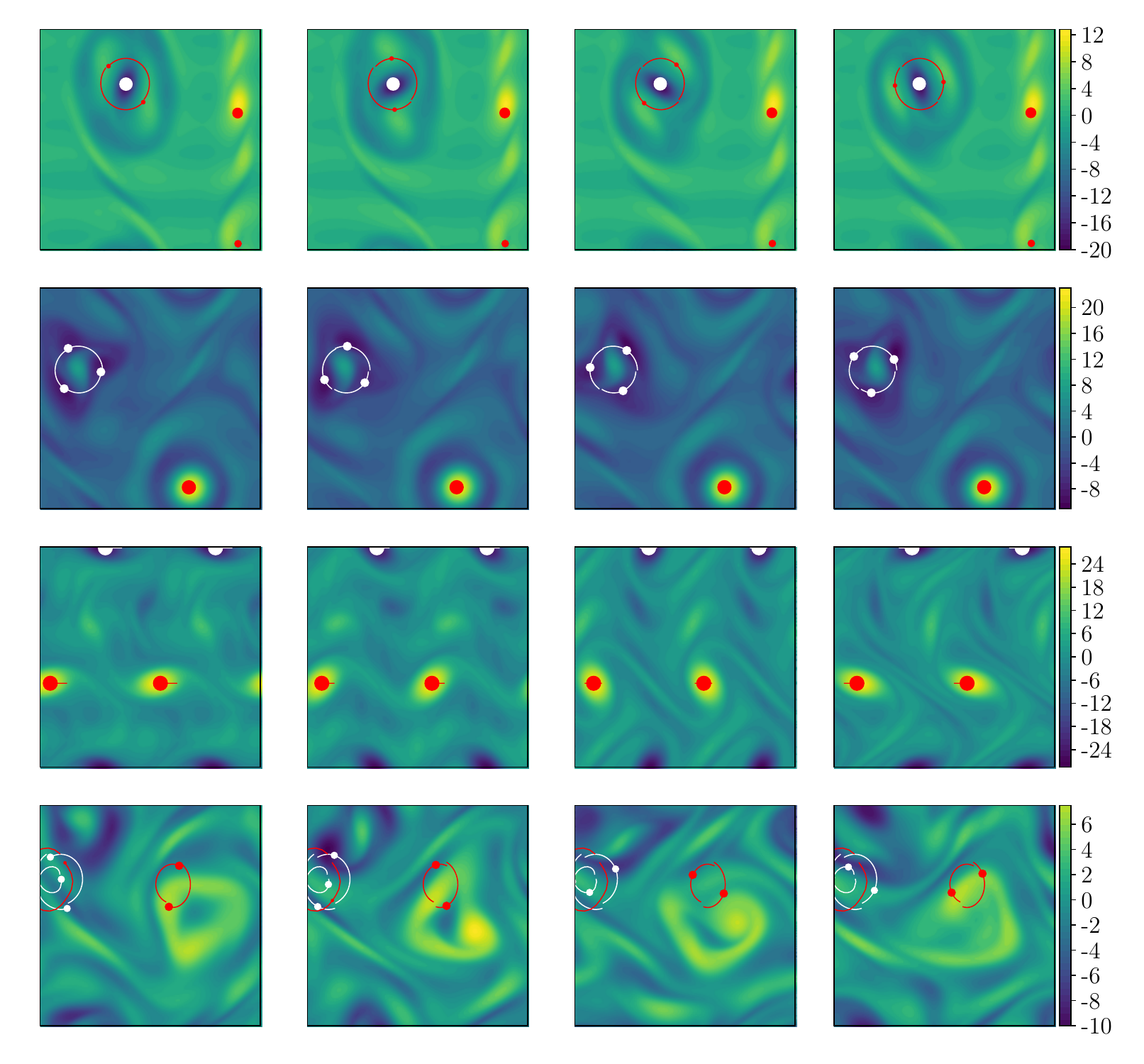}
    \caption{
    Out-of-plane vorticity (contours) of three RPOs from Kolmogorov flow at $\Rey = 100$.
    Snapshots in each row are sampled uniformly in time along the period of that solution. Overlaid on each snapshot are the point vortices at the corresponding (proportional) time along the period of the matched point vortex RPO. The size of the markers is proportional to the circulation of the vortex, and the lines denote the path swept out by each point vortex over its full evolution. Red (white) markers denote positive (negative) circulations. The reference RPO period $T_{\text{RPO}}$, reference RPO shift $s_{\text{RPO}}$, point vortex RPO period $T$ and point vortex shifts $\bm s$ for each solution are: (Top row) $T_{\text{RPO}} = 1.34$, $s_{\text{RPO}} = 0.0075$, $T = 1.2$, $\bm s = (-0.032, 0.012)$; (Second row) $T_{\text{RPO}} = 1.79$, $s_{\text{RPO}} = 0.085$, $T = 1.755$, $\bm s = (0.05, 0.006)$; (Third row) $T_{\text{RPO}} = 1.27$, $s_{\text{RPO}} = 0.69$, $T = 1.27$, $\bm s = (0.48, 0)$. This point vortex solution is a vortex crystal (relative equilibrium); (Bottom row) $T_{\text{RPO}} = 4.16$, $s_{\text{RPO}} = -0.756$, $T = 4.34$, $\bm s = (0.04, -0.12)$. 
    }
    \label{fig:upo_animations}
\end{figure}

A number of different experiments were run, varying the circulation threshold $\varepsilon$ controlling which vortices are retained, and the annealing parameter $\kappa$ in equation (\ref{eq:total_loss}). 
These experiments are summarised in table \ref{tab:upo_experiments}, while examples of some converged solutions, overlaid over the reference Kolmogorov RPO, are shown in figure \ref{fig:upo_animations}. 
The labelling procedure was initialised with the 252 RPOs converged at $\Rey = 100$ (151 solutions from the original library, 101 new solutions from the continuation).
Three different initialisations were run for each of these reference turbulent RPOs, by extracting the $N_v-1, \, N_v$ and $N_v + 1$ most dynamically important vortex cores over the period of the reference RPO (the overbar indicating the modal number of point vortices is dropped from now on).
\JP{Vortex numbers are typically $2 \leq N_v \leq 8$, with most periodic orbits being well described with $N_v = 4$ or 5.}
The `Total Convergences' column in table \ref{tab:upo_experiments} reports the total number of convergences (larger than \REV{252} for each experiment due to the three initialisations run for each RPO), \JP{while} the `Total Matches' column reports how many of the reference RPOs yielded at least one converged point vortex RPO.

Overall, table \ref{tab:upo_experiments} indicates that all parameter settings achieve very large numbers of converged solutions, though these convergences may in some cases differ substantially from the reference simulation --
\JP{point vortex states which are `similar' to the underlying turbulent periodic orbit are quantified by a dynamical relevance metric $\mathscr L_{\text{rel}} \leq 1$ which is introduced and discussed below.}
In some cases in table \ref{tab:upo_experiments} the parameter $\kappa$ is labelled `Anneal': we use this terminology to describe an optimisation which is initialised with $\kappa = 1$, and in which $\kappa$ is then decreased by $1/N_{\text{opt}}$ at each optimisation step.
This annealing aims to initially closely match the reference turbulent RPO, while slowly increasing the relative importance of the $\mathscr L_{\text{RPO}}$ throughout the optimisation procedure. 
This scheduling results in a slight increase in the total number of point vortex RPO convergences compared to a constant equal weighting ($\kappa = 0.5$) of $\mathscr L_{\text{RPO}}$ and $\mathscr L_{\text{match}}$.
Ignoring the $\mathscr L_{\text{match}}$ term (setting $\kappa = 1$) results in a significant increase in the total number of point vortex RPO convergences.


The solutions reported in the top three rows of figure \ref{fig:upo_animations} all represent cases where the converged point vortex dynamics closely track the dominant vortex cores in the original solution. 
They include cases featuring a multipolar structure (top row) and a bound state or triangular vortex (middle row) -- in both of these examples the point vortex RPO involves a permutation of vortices at each period. 
The final example is a crystal-like structure in which two rows of same-signed vortices 
translate to the right.
In all three cases the period of the converged point vortex solution matches the reference turbulent RPO to within a few percent or better. 

However, by visual inspection the dynamics of some converged point vortex RPOs differ qualitatively from the reference turbulent RPO. 
An example of which is shown in the bottom row of figure \ref{fig:upo_animations}, in which a six vortex system has been converged which only slightly resembles the reference RPO -- there are two trapped systems of $N_v=4$ and $N_v=2$ vortices which roughly align with large scale vortices in the turbulent flow.

To quantify the dynamical relevance of the point vortex state to the original turbulent dynamics, we compute the translation-and-time-reduced metric:
\begin{equation}
    \mathscr L_{\text{rel}} :=  \min_{\bm{s}, \Delta t}  \frac{1}{N_{S}} \sum_{j = 1}^{N_{S}} \frac{\| G(\bm{x}; \mathcal{T}^{\bm s} \mathbf f^{t_j^{v} + \Delta t} (\mathbf x^*), \bm{\Gamma}) - \sum_{i} \ind_i (\bm{x},t_j) \omega(\bm{x}, t_j)  \|^2}{\| \sum_{i} \ind_i (\bm{x},t_j) \omega(\bm{x},t_j) \|^2},
    \label{eq:l_rel}
\end{equation}
where the indicator function $\ind_i (\bm{x},t) = 1$ if $\bm{x} \in \mathcal{V}_{i}$ at time $t$ -- i.e. we are comparing only the vortex cores in the original RPO to the point vortex reconstruction. 
The solutions are compared at equally spaced times $t_j^{v} := j T / N_{S}$ and $t_j = j T_{\text{RPO}} / N_{S}$ along the periods of point vortex RPO $T$ and the reference turbulent RPO $T_{\text{RPO}}$, respectively.
Due to potential small mismatches between the alignment of the vortices in the two solutions in space and time, we search in (\ref{eq:l_rel}) over a restricted range of (constant) temporal and spatial shifts: $\Delta t \in
\{0, \pm 10 \delta t, \pm 25\delta t, \pm 50\delta t, \pm 100\delta t\}$, and
$s_x, s_y \in
\{0, \pm L/100, \pm L/40, \pm L/20, \pm L/10\}$.

It is worth considering what a `good' value of $\mathscr L_{\text{rel}}$ would be. 
For unrelated point vortex observables and turbulent RPOs we find $\mathscr L_{\text{rel}} \approx 2$ (the expected result given we are comparing two near-orthogonal vectors in a high-dimensional space).
For the dynamically similar states obtained in figure \ref{fig:upo_animations} we obtain values of
$\mathscr L_{\text{rel}} = 0.47, 0.49, 0.46$ for the first, second and third rows respectively.
The state in the bottom row of figure \ref{fig:upo_animations} only achieves a value of $\mathscr L_{\text{rel}} = 1.58$, indicating a poor representation of the turbulent dynamics.
The relatively `large' values of $\mathscr L_{\text{rel}}$ in the well-matched configurations can be rationalised by the fact that the smoothed point vortex observable is constructed as a superposition of Gaussians, with the shape of the original vortex core modelled by a variance which is proportional to the vortex area, $\sigma_{\alpha}^2 = 0.1 \mathcal{A}_{\alpha}$: we are comparing fields which can differ quite significantly even in the best scenarios. 
As such, we find that $\mathscr L_{\text{rel}}\leq 1$ is a sensible threshold to discern whether the converged point vortex RPO captures critical features of the original turbulent solution.
\JP{We also checked all point vortex states with $\mathscr L_{\text{rel}} \leq 1$ for uniqueness by rescaling variables such that $\max |\Gamma_j| = 1$, before comparing their rescaled periods and values of the rescaled Hamiltonian. 
All solutions are unique.
}

\begin{figure}
    \centering
        \includegraphics[width=0.7\linewidth]{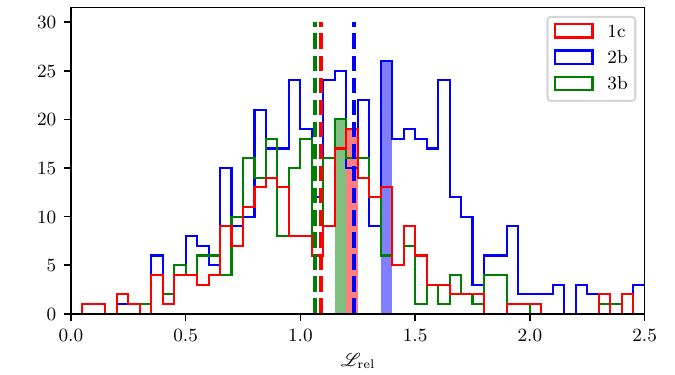}
    \caption{Histograms of the dynamical relevance metric (\ref{eq:l_rel}) for experiments 1c, 2b and 3b detailed in table \ref{tab:upo_experiments}, with a bin size of 0.05. The mean and mode of each experiment are indicated by the correspondingly coloured bold dashed line and filled bars, respectively. }
    \label{fig:all_exp_match_loss_both}
\end{figure}
A comparison of the distributions of $\mathscr L_{\text{rel}}$ across 3 representative experiments from the list in table \ref{tab:upo_experiments} (1c, 2b and 3b) is reported in figure \ref{fig:all_exp_match_loss_both}.
These point vortex RPO searches all enforced removal of vortices whose circulation shrunk in the optimisation below a threshold of $\varepsilon=0.05$, and are distinguished from one another by the relative weight placed on the two contributions to the loss (\ref{eq:total_loss}):
In run 1c the `RPO' contribution (\ref{eq:loss_upo}) was included alongside the matching constraint (\ref{eq:loss_match}), while it represented the entire loss in 2b (no matching term). In 3b annealing was applied to incrementally increase the contribution of the RPO term over the course of the optimisation. 
All of the search configurations summarised in figure \ref{fig:all_exp_match_loss_both} show a wide spread in $\mathscr{L}_{\text{rel}}$. Notably the RPO-only results (2b) show much higher $\mathscr L_{\text{rel}}$ when no requirement is made that the point vortices attempt to shadow the vortex cores.
We focus for the remainder of the section on the results of experiment 3b, which has the lowest mean (and mode) in $\mathscr L_{\text{rel}}$, though the results are representative of the suite of experiments in general.
Note that the value of the circulation cutoff, $\varepsilon$, selected for analysis here (the largest, $\varepsilon = 0.05$) leads to an increased number of convergences relative to calculations which retain larger numbers of weak vortices, though the overall distribution of results are not substantially changed from those shown in figure \ref{fig:all_exp_match_loss_both}.

\begin{figure}
    \centering
    \begin{minipage}{0.5\textwidth}
        \centering
        \includegraphics[width=0.9\linewidth]{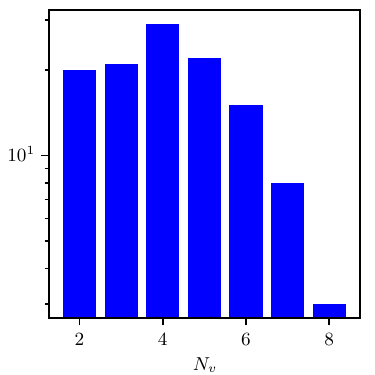}
    \end{minipage}%
    \begin{minipage}{0.5\textwidth}
        \centering
        \includegraphics[width=0.9\linewidth]{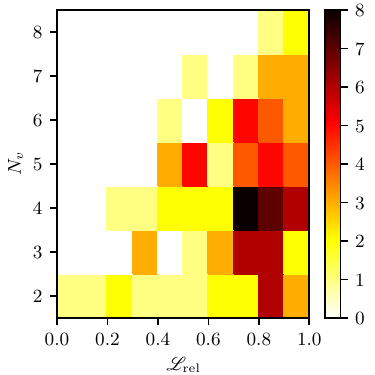}
    \end{minipage}
    \caption{(Left) The total number of point vortex RPO convergences at each value of $N_v$ for experiment 3b in table \ref{tab:upo_experiments} which satisfy $\mathscr L_{\text{rel}} < 1$. (Right) Histogram showing the distribution of $\mathscr L_{\text{rel}}$ for experiment 3b as a function of $N_v$. The bin size for $\mathscr L_{\text{rel}}$ is 0.1. }
    \label{fig:Nconv_match_loss_by_N_3b_delta01}
\end{figure}

\subsection{Discussion}

We now explore the properties of the library of converged point vortex states from experiment 3b and their relationship to the three classes of Kolmogorov RPO discussed in \S\ref{sec:continuation}.
The number of convergences from this experiment satisfying $\mathscr L_{\text{rel}} < 1$ is visualised as a function of the number of point vortices $N_v$ in the left panel of figure \ref{fig:Nconv_match_loss_by_N_3b_delta01}.
There is a preference for small numbers of vortices, $N_v \leq 5$, with the most solutions found for $N_v=4$.
Intriguingly, there is no clear dependence in the reconstruction error, $\mathscr L_{\text{rel}}$, on the number of vortices, $N_v$, other than an increase in the minimum observed value of $\mathscr L_{\text{rel}}$ with increasing $N_v$.
This is perhaps unexpected as intuitively one might think that placing more point vortices in the system would result in a more faithful representation of the turbulent dynamics. 
However, the solutions in the turbulent RPO library at $\Rey = 100$ tend to be dominated by a small number of vortex cores (see for example the first columns in figures \ref{fig:class2_eg} and \ref{fig:class3_eg}), and additional point vortices are attempting to represent weak, small-scale filamentary structures which are poorly represented with Gaussians. 

\begin{figure}
    \centering
    \includegraphics[width=\linewidth]{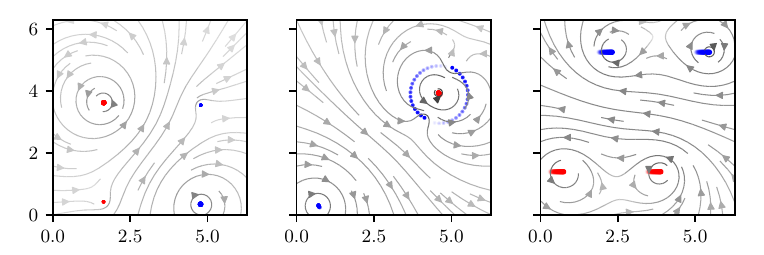}
    \caption{Evolution of three samples point vortex RPOs with $N_v = 4$. Trajectories are plotted as red (blue) points denoting positive (negative) circulation, such that increasing opacity denoting increasing time along the period and the separation of points along a trajectory gives an indication of the speed of that vortex. Streamlines of the final (most opaque) vortex configuration are also shown, such that darker streamlines denote greater local induced speed. Each system is simulated for one complete period. (Left) Vortex crystal configuration with $T = 2.16$, $\bm s = (-0.008, 0.002)$. (Middle) Tripolar configuration with $T = 1.75$, $\bm s = (-0.056, 0.073)$. Uniform crystal lattice configuration with $T = 1.27$, $\bm s = (0.48, 0)$.}
    \label{fig:pv_upo_egs}
\end{figure}

Some examples of solutions in the most common $N_v=4$ class of point vortex RPOs are reported in figure \ref{fig:pv_upo_egs}.
Most $N_v=4$ solutions appear as an approximately equispaced lattice, with two dominant vortices of opposite circulation. 
An example of one of these `quasi-crystal' configurations is shown in the left panel of figure \ref{fig:pv_upo_egs}. 
Other representative solutions for the $N_v = 4$ system are shown in the remaining two panels, and show a `tripole' structure and an oscillatory lattice, featuring rows of equal-signed vortices, respectively.

\begin{figure}
    \centering
    \begin{minipage}{0.5\textwidth}
        \centering
        \includegraphics[width=0.9\linewidth]{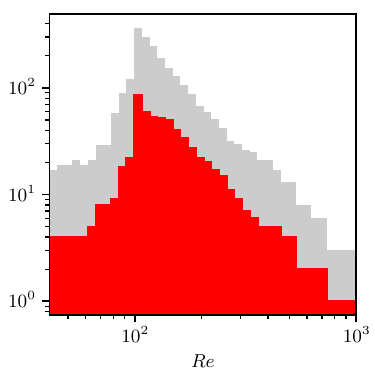}
    \end{minipage}%
    \begin{minipage}{0.5\textwidth}
        \centering
        \includegraphics[width=0.9\linewidth]{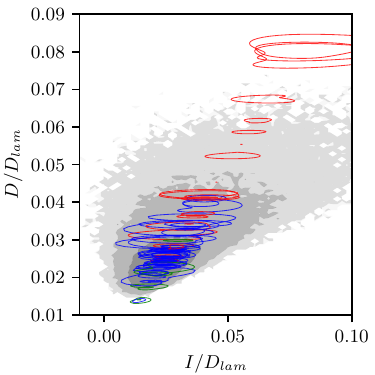}
    \end{minipage}
    \caption{Left panel: Histogram of all the distinct RPO branches which cross $\Rey = 100$ as a function of $\Rey$ in grey. The histogram of the branches which were successfully matched at $\Rey = 100$ is shown in red. 
    Right panel: Energy dissipation rate $D$ against energy production rate $I$ at $\Rey = 100$, both normalised by the laminar dissipation $D_{lam} = \Rey / (2n^2)$, for the turbulent RPOs which were successfully matched with a point vortex RPO. The RPOs are coloured according to their branch class; (1) red, (2) blue and (3) green. The grey background is the PDF computed from a trajectory of $10^5$ samples, separated by 1 advective time unit. The contour levels of the PDF are spaced logarithmically.}
    \label{fig:matched_both}
\end{figure}

A key question is whether the successful fitting of a point vortex solution to a Kolmogorov RPO says anything about the possible connection of the original RPO to a solution of the \JP{(unforced)} Euler equation. 
To explore this point we first examine how many of the subset of RPOs which were successfully labelled with a `representative' point vortex solution at $Re=100$ (quantified by $\mathscr L_{\text{rel}} <1$) can be continued to \REV{higher $Re$}.
This is done via the histogram in figure \ref{fig:matched_both}, which compares the number of labelled branches to the total number of solutions as $Re$ varies; both scale similarly with $Re$.

\begin{table}
  \begin{center}
    \def~{\hphantom{0}}
    \begin{tabular}{c|c|c|c|c}
          & \multicolumn{2}{c|}{$\Rey = 100$} & \multicolumn{2}{c}{$\Rey > 200$} \\ \hline
         RPO Class & Branches labelled  & \% labelled & Branches labelled  & \%  labelled \\ \hline
        1 & 21 &  33.3 \% &  14 &  43.75 \%    \\
        2 & 32 & 18.4 \% & 4 & 25 \%  \\
        3 & 7 & 46.7 \% & 6 & 50 \% \\ \hline
    \end{tabular}
    \caption{
    The number and percentage of successfully labelled ($\mathscr L_{\text{rel}} < 1$) branches for each of the three distinct classes of solution branches identified in section \ref{sec:continuation}.
    The $\Rey = 100$ columns report these figures for experiment 3b, run on the library of 252 RPOs at $\Rey = 100$. 
    The $\Rey > 200$ columns report these figures for the same labelling procedure, but the reference solutions considered are instead all the terminal branch solutions which were continued to above $\Rey = 200$.
    }
    \label{tab:class_matches}
  \end{center}
\end{table}

More revealing is the right panel of figure \ref{fig:matched_both} which identifies point-vortex-labelled RPOs on a production/dissipation plot at $Re=100$, where the solutions are coloured according to the class assigned in \S \ref{sec:continuation}.
The figure appears to be dominated by solutions from classes 1 (red, dissipation stronger than turbulent PDF) and 2 (blue, dissipation within turbulent PDF).
However, this figure is distorted by the large number of solutions in class 2, while the labelling success rate as a percentage of the cardinality of the class was significantly larger for class (1) and class (3).
These statistics are summarised in table \ref{tab:class_matches}, which reports the number and percentage of successfully labelled solution branches (one branch may contain multiple solutions at a given $Re$) for each of the three distinct branch classes. 
The branches which could be continued the highest in $\Rey$ by the arclength continuation (class 3) were the most likely to be successfully labelled with a point vortex RPO (albeit with only a small sample size). 
The RPOs in class (1) were the second most likely to be successfully labelled. 

\begin{figure}
    \centering
    \includegraphics[width=\linewidth, trim={0 5cm 0 2cm},clip]{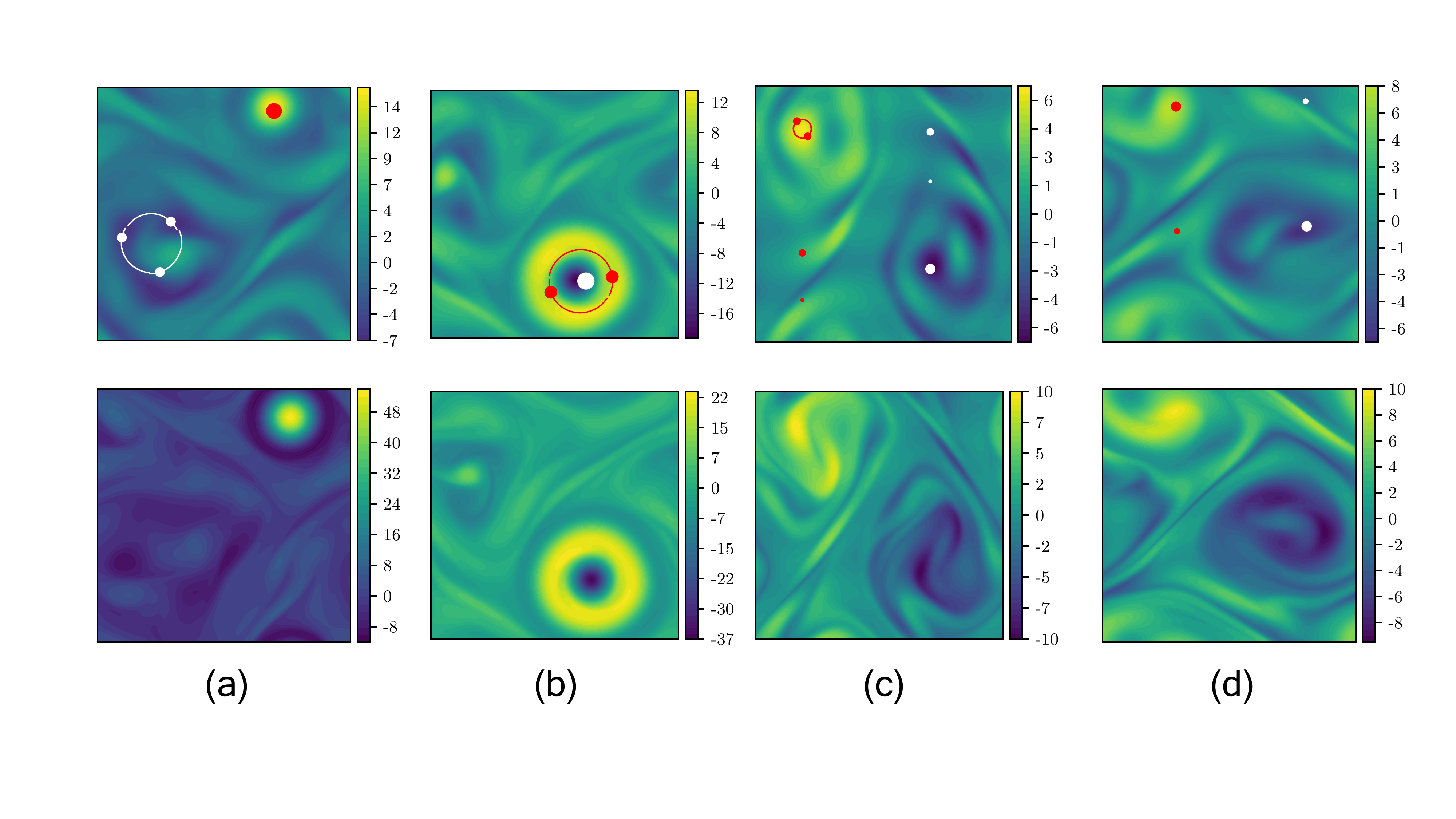}
    \caption{
    Contours of vorticity for four RPOs in Kolmogorov flow (left to right).
    Top row shows a snapshot at $Re=100$ along with the point vortex RPO overlaid as in previous figures. 
    The bottom row is the final state obtained after arclength continuation up in $Re$.
    The two leftmost RPOs belong to `class 1', while the two rightmost RPOs belong to `class 3' as described in \S\ref{sec:continuation}.
    The reference RPO period and shift, $T_{\text{RPO}}$, shift $s_{\text{RPO}}$, point vortex RPO period $T$ and shifts $\bm{s}$ for each labelled solution on the top row are: 
    (a) $T_{\text{RPO}} = 2.99$, $s_{\text{RPO}} = -0.44$, $T = 2.97$, $\bm s = (-0.041, -0.023)$;
    (b) $T_{\text{RPO}} = 1.97$, $s_{\text{RPO}} =  -0.43$, $T = 1.89$, $\bm s = (0.01, -0.01)$;
    (c) $T_{\text{RPO}} = 4.61$, $s_{\text{RPO}} =  0.071$, $T = 4.695$, $\bm s = (-0.008, -0.001)$;
    (d) $T_{\text{RPO}} = 4.67$, $s_{\text{RPO}} =  -0.17$, $T = 4.497$, $\bm s = (0.016, -0.018)$. The point vortex solution (d) is a travelling wave.
    The final solutions on the bottom row were converged at $\Rey = 185.69, 175.003, 345.297$ and $277.23$, from (a) to (d) respectively.
    }
    \label{fig:snapshots}
\end{figure}

Two examples from both classes (1 and 3) are reported in figure \ref{fig:snapshots} with the point vortex solution overlaid, along with a higher-$Re$ state from the same solution branch. 
In the cases from class 1, the point vortex periodic orbits correspond to features of the turbulent solution which remain as $Re$ grows.
For instance, note the multipole structure in the second column approximating ring of vorticity, which amplifies under increasing $Re$, while the weaker triangular vortex in the same snapshot is not represented in the point vortex solution, and weakens as $Re$ increases. 
Similar effects are also apparent in the earlier examples in figure \ref{fig:upo_animations} from class 1 (top 3 rows of that figure).

On the other hand, the solutions in class 3 have resulted in the initialisation of weaker point vortices on small filamentary regions (e.g. see the weaker pair of vortices in the second class 3 example in figure \ref{fig:snapshots}). 
These structures at $Re=100$ are a feature of many Kolmogorov RPOs and are tied to the $n=4$ wavenumber of the forcing \citep{Chandler2013,lucas2015}.
Continuation up in $Re$ changes these features substantially, where they would no longer be identified by the vortex `extraction' algorithm (\S\ref{sec:vort_id}) used to initialise the point vortex positions. 
In these cases, it appears we have found a point vortex RPO which can mimic the weaker viscous dynamics, without implying anything stronger about the behaviour as $Re\to \infty$.

\JP{
The identification by the point vortex fit of structures in class 1 which persist as $Re$ increases is consistent with the connection of these RPOs to exact solutions of the Euler equation as $Re\to \infty$, as suggested by their asymptotic scalings.
Under continuation, these RPOs tend to be increasingly dominated by large-scale, coherent vortices while weaker filamentary structures lessen. 
Hence, the point vortex fit identifies large-scale flow features which are retained in the high $Re$ limit. 
} 
Interestingly, these class 1 solutions contain examples with many interacting vortices \citep[typical of high dissipation states seen in past work][]{Chandler2013,Page2024,Redfern2024} but also examples of condensate-like structures dominated by a pair of opposite signed vortices, which are reminiscent of the Euler solutions discussed in previous work exploring Euler/Navier-Stokes connections \citep{Kim2015, Kim2017, Zhigunov2023}.
\JP{
The weaker performance of the point vortex fit in classes 2 and 3, and failure to extract structure retained as $Re$ increases, reflects the fact that – particularly in class 2 – the flow field is dominated by a large scale vortex pair and thin filamentary structures. 
Trivially one can fit to the structure associated with the condensate (a traveling wave always results with $N_v=2$) but not the smaller-scale asymmetric features. 
}

\section{Conclusion}

This study presented a continuation effort of a large number of numerically exact recurrent solutions in two-dimensional Kolmogorov flow, over the range $Re \in (30, 1000)$. 
Libraries of Kolmogorov RPOs at different $\Rey$ were constructed from sets of `starting' solutions at $Re=40$ and $Re=100$, and used to reconstruct summary statistics of the turbulence.
The reconstruction was performed with a data-driven approach in which the weights in an RPO-expansion were determined from a fit to a single PDF.
The quality of the reconstruction deteriorated dramatically with changing $Re$, \JP{while the numerically assigned weights become poorly correlated with the instability of the underlying RPOs}.
This raises interesting problems for the application of RPOs to studying turbulence, where a key motivation for the computational effort spent identifying and converging solutions is the ability to probe higher $Re$ with arclength continuation. 
Further work should explore if there are more `robust', perhaps longer-period, solutions with high dissipation excursions which can be used to say something about bursting events via continuation.
It also remains to be seen whether these conclusions are directly relevant to three-dimensional, wall-bounded flows.

The dependence of dissipation rate on $Re$ was used to delineate the large number of solution branches into three classes:
Class (1) RPOs  were characterised by a dissipation rate scaling $D \sim Re$, \JP{and an analysis of the scaling of flow variables indicated a direct connection to solutions of the unforced Euler equation as $Re\to \infty$.}
\JP{The RPOs on these branches feature} strong vortical structures that appeared largely decoupled from the specifics of the Kolmogorov forcing, \JP{and generalise the Euler connection to states beyond the `unimodal' structures considered previously \citep{Kim2015,Kim2017}}. 
Two other classes of solutions were identified which scaled either with the average turbulent dissipation (class (2), $D\sim Re^{-1/2}$) or weaker than it (class (3), $D \sim Re^{-1}$).
Class (2) solutions remained contained within a finite-$Re$ region, \REV{many repeatedly folding back and forth under} continuation.
\JP{Class (3) solutions show a scaling consistent with a connection to a forced Euler equation in the high-$Re$ limit}.

\JP{Motivated by the connection of large numbers of RPOs to inviscid solutions,}
a method was introduced to search for point vortex relative periodic orbits that reproduced the dominant vortex interactions in the original solution library at $Re=100$.
This was done via gradient-based minimisation of a scalar loss function. 
The method proved successful for a wide range of RPOs, but only for solutions in class (1) did it identify vortical structures which persist under increasing $Re$. 
\JP{The success can perhaps be attributed to the fact that these states become dominated by (smooth) large-scale structures as $Re$ increases at the expense of small scale filamentary vortices.
The success of the fit is consistent with a connection to exact inviscid solutions.
}

\vspace{0.5cm}
\noindent
\textbf{Declaration of Interests.} The authors report no conflict of interest.

\vspace{0.5cm}
\noindent
\textbf{Acknowledgements.} We are grateful to the referees for their detailed comments which have helped substantially to clarify the messages in this paper. This research has been supported by the UK Engineering and Physical Sciences Research Council through the MAC-MIGS Centre for Doctoral Training (EP/S023291/1). 
This work has made use of the resources provided by the Edinburgh Compute and Data Facility (ECDF) (http://www.ecdf.ed.ac.uk/). 
JP acknowledges support from UKRI Frontier Guarantee Grant EP/Y004094/1.

\bibliographystyle{jfm}
\bibliography{bib}

\begin{thebibliography}{69}
\expandafter\ifx\csname natexlab\endcsname\relax\def\natexlab#1{#1}\fi
\def\au#1{#1} \def\ed#1{#1} \def\yr#1{#1}\def\at#1{#1}\def\jt#1{\textit{#1}} \def\bt#1{#1}\def\bvol#1{\textbf{#1}} \def\vol#1{#1} \def\pg#1{#1} \def\publ#1{#1}\def\arxiv#1{#1}\def\org#1{#1}\def\st#1{\textit{#1}}

\bibitem[Aref \& Buren(2005)]{aref2005}
{\sc \au{Aref, Hassan} \& \au{Buren, Martin}} \yr{2005}  \at{Vortex triple rings}.  \jt{Physics of Fluids}  \bvol{17}.

\bibitem[Aref {\em et~al.\/}(2003)Aref, Newton, Stremler, Tokieda \& Vainchtein]{Aref2003}
{\sc \au{Aref, Hassan}, \au{Newton, Paul~K.}, \au{Stremler, Mark~A.}, \au{Tokieda, Tadashi} \& \au{Vainchtein, Dmitri~L.}} \yr{2003}  \at{Vortex crystals}.  \bt{In {\em Advances in Applied Mechanics\/}},  \pg{pp. 1--79}.  \publ{Elsevier}.

\bibitem[Aref \& Pomphrey(1982)]{Aref1982}
{\sc \au{Aref, H.} \& \au{Pomphrey, N.}} \yr{1982}  \at{Integrable and chaotic motions of four vortices i. the case of identical vortices}.  \jt{Proceedings of the Royal Society of London. Series A, Mathematical and Physical Sciences}  \bvol{380}~(1779),  \pg{359--387}.

\bibitem[Artuso {\em et~al.\/}(1990{\natexlab{{\em a\/}}})Artuso, Aurell \& Cvitanovic]{Artuso1990a}
{\sc \au{Artuso, R}, \au{Aurell, E} \& \au{Cvitanovic, P}} \yr{1990{\natexlab{{\em a\/}}}}  \at{Recycling of strange sets: I. cycle expansions}.  \jt{Nonlinearity}  \bvol{3}~(2),  \pg{325}.

\bibitem[Artuso {\em et~al.\/}(1990{\natexlab{{\em b\/}}})Artuso, Aurell \& Cvitanovic]{Artuso_1990b}
{\sc \au{Artuso, R}, \au{Aurell, E} \& \au{Cvitanovic, P}} \yr{1990{\natexlab{{\em b\/}}}}  \at{Recycling of strange sets: Ii. applications}.  \jt{Nonlinearity}  \bvol{3}~(2),  \pg{361}.

\bibitem[Batchelor(1967)]{batchelor1967}
{\sc \au{Batchelor, G.K.}} \yr{1967} {\em An Introduction to Fluid Dynamics\/}.  \publ{Cambridge University Press}.

\bibitem[Batchelor(1956)]{Batchelor1956}
{\sc \au{Batchelor, G.~K.}} \yr{1956}  \at{On steady laminar flow with closed streamlines at large reynolds number}.  \jt{Journal of Fluid Mechanics}  \bvol{1}~(2),  \pg{177–190}.

\bibitem[Benzi {\em et~al.\/}(1992)Benzi, Colella, Briscolini \& Santangelo]{benzi1992}
{\sc \au{Benzi, R.}, \au{Colella, M.}, \au{Briscolini, M.} \& \au{Santangelo, P.}} \yr{1992}  \at{{A simple point vortex model for two‐dimensional decaying turbulence}}.  \jt{Physics of Fluids A: Fluid Dynamics}  \bvol{4}~(5),  \pg{1036--1039}.

\bibitem[Benzi {\em et~al.\/}(1987)Benzi, Patarnello \& Santangelo]{Benzi1987}
{\sc \au{Benzi, R.}, \au{Patarnello, S.} \& \au{Santangelo, P.}} \yr{1987}  \at{On the statistical properties of two-dimensional decaying turbulence}.  \jt{Europhysics Letters}  \bvol{3}~(7),  \pg{811}.

\bibitem[Bradbury {\em et~al.\/}(2018)Bradbury, Frostig, Hawkins, Johnson, Leary, Maclaurin, Necula, Paszke, Vander{P}las, Wanderman-{M}ilne \& Zhang]{jax2018github}
{\sc \au{Bradbury, James}, \au{Frostig, Roy}, \au{Hawkins, Peter}, \au{Johnson, Matthew~James}, \au{Leary, Chris}, \au{Maclaurin, Dougal}, \au{Necula, George}, \au{Paszke, Adam}, \au{Vander{P}las, Jake}, \au{Wanderman-{M}ilne, Skye} \& \au{Zhang, Qiao}} \yr{2018} {JAX}: composable transformations of {P}ython+{N}um{P}y programs.

\bibitem[Budanur {\em et~al.\/}(2017)Budanur, Short, Farazmand, Willis \& Cvitanović]{Budanur2017}
{\sc \au{Budanur, N.~B.}, \au{Short, K.~Y.}, \au{Farazmand, M.}, \au{Willis, A.~P.} \& \au{Cvitanović, P.}} \yr{2017}  \at{Relative periodic orbits form the backbone of turbulent pipe flow}.  \jt{Journal of Fluid Mechanics}  \bvol{833},  \pg{274–301}.

\bibitem[Campbell \& Ziff(1979)]{campbell1979vortex}
{\sc \au{Campbell, Laurence~J} \& \au{Ziff, Robert~M}} \yr{1979}  \at{Vortex patterns and energies in a rotating superfluid}.  \jt{Physical Review B}  \bvol{20}~(5),  \pg{1886}.

\bibitem[Carnevale \& Kloosterziel(1994)]{Carnevale1994}
{\sc \au{Carnevale, G.~F.} \& \au{Kloosterziel, R.~C.}} \yr{1994}  \at{Emergence and evolution of triangular vortices}.  \jt{Journal of Fluid Mechanics}  \bvol{259},  \pg{305–331}.

\bibitem[Carton {\em et~al.\/}(1989)Carton, Flierl \& Polvani]{carton1989}
{\sc \au{Carton, X.~J.}, \au{Flierl, G.~R.} \& \au{Polvani, L.~M.}} \yr{1989}  \at{The generation of tripoles from unstable axisymmetric isolated vortex structures}.  \jt{Europhysics Letters}  \bvol{9}~(4),  \pg{339}.

\bibitem[Chandler \& Kerswell(2013)]{Chandler2013}
{\sc \au{Chandler, G.~J.} \& \au{Kerswell, R.~R.}} \yr{2013}  \at{Invariant recurrent solutions embedded in a turbulent two-dimensional {K}olmogorov flow}.  \jt{Journal of Fluid Mechanics}  \bvol{722},  \pg{554–595}.

\bibitem[Christiansen {\em et~al.\/}(1997)Christiansen, Cvitanovic \& Putkaradze]{Christiansen1997}
{\sc \au{Christiansen, F}, \au{Cvitanovic, P} \& \au{Putkaradze, V}} \yr{1997}  \at{Spatiotemporal chaos in terms of unstable recurrent patterns}.  \jt{Nonlinearity}  \bvol{10}~(1),  \pg{55}.

\bibitem[Cleary \& Page(2023)]{Cleary2023}
{\sc \au{Cleary, Andrew} \& \au{Page, Jacob}} \yr{2023}  \at{{Exploring the free-energy landscape of a rotating superfluid}}.  \jt{Chaos: An Interdisciplinary Journal of Nonlinear Science}  \bvol{33}~(10),  \pg{103123}.

\bibitem[Crouse(2016)]{crouse2016}
{\sc \au{Crouse, David~F.}} \yr{2016}  \at{On implementing 2d rectangular assignment algorithms}.  \jt{IEEE Transactions on Aerospace and Electronic Systems}  \bvol{52}~(4),  \pg{1679--1696}.

\bibitem[Crowdy(1999)]{Crowdy1999}
{\sc \au{Crowdy, Darren}} \yr{1999}  \at{{A class of exact multipolar vortices}}.  \jt{Physics of Fluids}  \bvol{11}~(9),  \pg{2556--2564}.

\bibitem[Crowdy(2003)]{Crowdy2003a}
{\sc \au{Crowdy, Darren~G.}} \yr{2003}  \at{Stuart vortices on a sphere}.  \jt{Journal of Fluid Mechanics}  \bvol{498},  \pg{381 -- 402}.

\bibitem[Cvitanovi{\'c} {\em et~al.\/}(2016)Cvitanovi{\'c}, Artuso, Mainieri, Tanner \& Vattay]{ChaosBook}
{\sc \au{Cvitanovi{\'c}, P.}, \au{Artuso, R.}, \au{Mainieri, R.}, \au{Tanner, G.} \& \au{Vattay, G.}} \yr{2016} {\em Chaos: Classical and Quantum\/}.  \publ{Copenhagen: Niels Bohr Inst.}

\bibitem[Cvitanović(1991)]{CVITANOVIC1991}
{\sc \au{Cvitanović, Predrag}} \yr{1991}  \at{Periodic orbits as the skeleton of classical and quantum chaos}.  \jt{Physica D: Nonlinear Phenomena}  \bvol{51}~(1),  \pg{138--151}.

\bibitem[Cvitanović(2013)]{cvitanovic2013}
{\sc \au{Cvitanović, Predrag}} \yr{2013}  \at{Recurrent flows: the clockwork behind turbulence}.  \jt{Journal of Fluid Mechanics}  \bvol{726},  \pg{1–4}.

\bibitem[Dresdner {\em et~al.\/}(2022)Dresdner, Kochkov, Norgaard, Zepeda-Núñez, Smith, Brenner \& Hoyer]{LCspectral}
{\sc \au{Dresdner, Gideon}, \au{Kochkov, Dmitrii}, \au{Norgaard, Peter}, \au{Zepeda-Núñez, Leonardo}, \au{Smith, Jamie~A.}, \au{Brenner, Michael~P.} \& \au{Hoyer, Stephan}} \yr{2022} Learning to correct spectral methods for simulating turbulent flows.

\bibitem[Faisst \& Eckhardt(2003)]{Faisst2003}
{\sc \au{Faisst, Holger} \& \au{Eckhardt, Bruno}} \yr{2003}  \at{Traveling waves in pipe flow}.  \jt{Phys. Rev. Lett.}  \bvol{91},  \pg{224502}.

\bibitem[Farazmand(2016)]{Farazmand2016}
{\sc \au{Farazmand, Mohammad}} \yr{2016}  \at{An adjoint-based approach for finding invariant solutions of navier–stokes equations}.  \jt{Journal of Fluid Mechanics}  \bvol{795},  \pg{278 -- 312}.

\bibitem[Gallet \& Young(2013)]{Gallet2013}
{\sc \au{Gallet, Basile} \& \au{Young, William~R.}} \yr{2013}  \at{A two-dimensional vortex condensate at high reynolds number}.  \jt{Journal of Fluid Mechanics}  \bvol{715},  \pg{359–388}.

\bibitem[Gibson {\em et~al.\/}(2008)Gibson, Halcrow \& Cvitanovic]{Gibson2008}
{\sc \au{Gibson, J.~F.}, \au{Halcrow, J.} \& \au{Cvitanovic, P.}} \yr{2008}  \at{Visualizing the geometry of state space in plane {C}ouette flow}.  \jt{Journal of Fluid Mechanics}  \bvol{611},  \pg{107–130}.

\bibitem[Graham \& Floryan(2021)]{Graham2021}
{\sc \au{Graham, Michael~D.} \& \au{Floryan, Daniel}} \yr{2021}  \at{Exact coherent states and the nonlinear dynamics of wall-bounded turbulent flows}.  \jt{Annual Review of Fluid Mechanics}  \bvol{53}~(1),  \pg{227--253}.

\bibitem[Hall \& Sherwin(2010)]{hall_sherwin_2010}
{\sc \au{Hall, Philip} \& \au{Sherwin, Spencer}} \yr{2010}  \at{Streamwise vortices in shear flows: harbingers of transition and the skeleton of coherent structures}.  \jt{Journal of Fluid Mechanics}  \bvol{661},  \pg{178–205}.

\bibitem[Hall \& Smith(1991)]{Hall_Smith_1991}
{\sc \au{Hall, P.} \& \au{Smith, F.~T.}} \yr{1991}  \at{On strongly nonlinear vortex/wave interactions in boundary-layer transition}.  \jt{Journal of Fluid Mechanics}  \bvol{227},  \pg{641–666}.

\bibitem[Hof {\em et~al.\/}(2004)Hof, Van~Doorne, Westerweel, Nieuwstadt, Faisst, Eckhardt, Wedin, Kerswell \& Waleffe]{hof2004}
{\sc \au{Hof, Bjorn}, \au{Van~Doorne, Casimir~WH}, \au{Westerweel, Jerry}, \au{Nieuwstadt, Frans~TM}, \au{Faisst, Holger}, \au{Eckhardt, Bruno}, \au{Wedin, Hakan}, \au{Kerswell, Richard~R} \& \au{Waleffe, Fabian}} \yr{2004}  \at{Experimental observation of nonlinear traveling waves in turbulent pipe flow}.  \jt{Science}  \bvol{305}~(5690),  \pg{1594--1598}.

\bibitem[Jiménez(2020)]{jimenez_2020}
{\sc \au{Jiménez, Javier}} \yr{2020}  \at{Dipoles and streams in two-dimensional turbulence}.  \jt{Journal of Fluid Mechanics}  \bvol{904},  \pg{A39}.

\bibitem[Kawahara \& Kida(2001)]{Kawahara2001}
{\sc \au{Kawahara, G.} \& \au{Kida, S.}} \yr{2001}  \at{Periodic motion embedded in plane {C}ouette turbulence: regeneration cycle and burst}.  \jt{Journal of Fluid Mechanics}  \bvol{449},  \pg{291–300}.

\bibitem[Kawahara {\em et~al.\/}(2012)Kawahara, Uhlmann \& van Veen]{Kawahara2012}
{\sc \au{Kawahara, G.}, \au{Uhlmann, M.} \& \au{van Veen, L.}} \yr{2012}  \at{The significance of simple invariant solutions in turbulent flows}.  \jt{Annual Review of Fluid Mechanics}  \bvol{44}~(1),  \pg{203--225}.

\bibitem[Kerswell(2005)]{Kerswell2005}
{\sc \au{Kerswell, R~R}} \yr{2005}  \at{Recent progress in understanding the transition to turbulence in a pipe}.  \jt{Nonlinearity}  \bvol{18}~(6),  \pg{R17}.

\bibitem[Kim {\em et~al.\/}(2017)Kim, Miyaji \& Okamoto]{Kim2017}
{\sc \au{Kim, Sun-Chul}, \au{Miyaji, Tomoyuki} \& \au{Okamoto, Hisashi}} \yr{2017}  \at{Unimodal patterns appearing in the two-dimensional navier–stokes flows under general forcing at large reynolds numbers}.  \jt{European Journal of Mechanics - B/Fluids}  \bvol{65},  \pg{234--246}.

\bibitem[Kim \& Okamoto(2015)]{Kim2015}
{\sc \au{Kim, Sun-Chul} \& \au{Okamoto, Hisashi}} \yr{2015}  \at{Unimodal patterns appearing in the kolmogorov flows at large reynolds numbers}.  \jt{Nonlinearity}  \bvol{28}~(9),  \pg{3219}.

\bibitem[Kingma \& Ba(2015)]{Kingma2015}
{\sc \au{Kingma, Diederik~P.} \& \au{Ba, Jimmy}} \yr{2015} Adam: {A} method for stochastic optimization.  \bt{In {\em 3rd International Conference on Learning Representations, {ICLR} 2015, San Diego, CA, USA, May 7-9, 2015, Conference Track Proceedings\/} (ed. \ed{Yoshua Bengio \& Yann LeCun})}.

\bibitem[Kochkov {\em et~al.\/}(2021)Kochkov, Smith, Alieva, Wang, Brenner \& Hoyer]{Kochkov2021}
{\sc \au{Kochkov, Dmitrii}, \au{Smith, Jamie~A.}, \au{Alieva, Ayya}, \au{Wang, Qing}, \au{Brenner, Michael~P.} \& \au{Hoyer, Stephan}} \yr{2021}  \at{Machine learning{\textendash}accelerated computational fluid dynamics}.  \jt{Proceedings of the National Academy of Sciences}  \bvol{118}~(21).

\bibitem[Krygier {\em et~al.\/}(2021)Krygier, Pughe-Sanford \& Grigoriev]{Krygier2021}
{\sc \au{Krygier, Michael~C.}, \au{Pughe-Sanford, Joshua~L.} \& \au{Grigoriev, Roman~O.}} \yr{2021}  \at{Exact coherent structures and shadowing in turbulent taylor–couette flow}.  \jt{Journal of Fluid Mechanics}  \bvol{923},  \pg{A7}.

\bibitem[Kuhn(1955)]{kuhn1955}
{\sc \au{Kuhn, H.~W.}} \yr{1955}  \at{The hungarian method for the assignment problem}.  \jt{Naval Research Logistics Quarterly}  \bvol{2}~(1-2),  \pg{83--97}.

\bibitem[{Lewis} \& {Ratiu}(1996)]{lewis1996}
{\sc \au{{Lewis}, D.} \& \au{{Ratiu}, T.}} \yr{1996}  \at{{Rotating n-gon/ kn-gon vortex configurations}}.  \jt{Journal of Nonlinear Science}  \bvol{6}~(5),  \pg{385--414}.

\bibitem[Lucas \& Kerswell(2017)]{lucas2017}
{\sc \au{Lucas, Dan} \& \au{Kerswell, Rich}} \yr{2017}  \at{Sustaining processes from recurrent flows in body-forced turbulence}.  \jt{Journal of Fluid Mechanics}  \bvol{817},  \pg{R3}.

\bibitem[Lucas \& Kerswell(2015)]{lucas2015}
{\sc \au{Lucas, Dan} \& \au{Kerswell, Rich~R.}} \yr{2015}  \at{Recurrent flow analysis in spatiotemporally chaotic 2-dimensional kolmogorov flow}.  \jt{Physics of Fluids}  \bvol{27}~(4),  \pg{045106}.

\bibitem[McCormack {\em et~al.\/}(2024)McCormack, Cavalieri \& Hwang]{McCormack2024}
{\sc \au{McCormack, Matthew}, \au{Cavalieri, André~V.G.} \& \au{Hwang, Yongyun}} \yr{2024}  \at{Multi-scale invariant solutions in plane couette flow: a reduced-order model approach}.  \jt{Journal of Fluid Mechanics}  \bvol{983}.

\bibitem[Montemuro {\em et~al.\/}(2020)Montemuro, White, Klewicki \& Chini]{Montemuro2020}
{\sc \au{Montemuro, Brandon}, \au{White, Christopher~M.}, \au{Klewicki, Joseph~C.} \& \au{Chini, Gregory~P.}} \yr{2020}  \at{A self-sustaining process theory for uniform momentum zones and internal shear layers in high reynolds number shear flows}.  \jt{Journal of Fluid Mechanics}  \bvol{901},  \pg{A28}.

\bibitem[Morel \& Carton(1994)]{Morel1994}
{\sc \au{Morel, Yves~G.} \& \au{Carton, Xavier~J.}} \yr{1994}  \at{Multipolar vortices in two-dimensional incompressible flows}.  \jt{Journal of Fluid Mechanics}  \bvol{267},  \pg{23–51}.

\bibitem[Okamoto(1994)]{Okamoto1994}
{\sc \au{Okamoto, H.}} \yr{1994}  \at{A variational problem arising in the two-dimensional navier-stokes equations with vanishing viscosity}.  \jt{Applied Mathematics Letters}  \bvol{7}~(1),  \pg{29--33}.

\bibitem[Page {\em et~al.\/}(2021)Page, Brenner \& Kerswell]{Page2021}
{\sc \au{Page, J.}, \au{Brenner, M.~P.} \& \au{Kerswell, R.~R.}} \yr{2021}  \at{Revealing the state space of turbulence using machine learning}.  \jt{Physical Review Fluids}  \bvol{6},  \pg{034402}.

\bibitem[Page {\em et~al.\/}(2024{\natexlab{{\em a\/}}})Page, Holey, Brenner \& Kerswell]{Page2024}
{\sc \au{Page, Jacob}, \au{Holey, Joe}, \au{Brenner, Michael~P.} \& \au{Kerswell, Rich~R.}} \yr{2024{\natexlab{{\em a\/}}}}  \at{Exact coherent structures in two-dimensional turbulence identified with convolutional autoencoders}.  \jt{Journal of Fluid Mechanics}  \bvol{991},  \pg{A10}.

\bibitem[Page {\em et~al.\/}(2024{\natexlab{{\em b\/}}})Page, Norgaard, Brenner \& Kerswell]{page2022recurrent}
{\sc \au{Page, Jacob}, \au{Norgaard, Peter}, \au{Brenner, Michael~P.} \& \au{Kerswell, Rich~R.}} \yr{2024{\natexlab{{\em b\/}}}}  \at{Recurrent flow patterns as a basis for two-dimensional turbulence: Predicting statistics from structures}.  \jt{Proceedings of the National Academy of Sciences}  \bvol{121}~(23),  \pg{e2320007121}.

\bibitem[Park \& Graham(2015)]{Park_Graham_2015}
{\sc \au{Park, Jae~Sung} \& \au{Graham, Michael~D.}} \yr{2015}  \at{Exact coherent states and connections to turbulent dynamics in minimal channel flow}.  \jt{Journal of Fluid Mechanics}  \bvol{782},  \pg{430–454}.

\bibitem[Redfern {\em et~al.\/}(2024)Redfern, Lazer \& Lucas]{Redfern2024}
{\sc \au{Redfern, Edward~M.}, \au{Lazer, Andrei~L.} \& \au{Lucas, Dan}} \yr{2024}  \at{Dynamically relevant recurrent flows obtained via a nonlinear recurrence function from two-dimensional turbulence}.  \jt{Phys. Rev. Fluids}  \bvol{9},  \pg{124401}.

\bibitem[Sakajo(2019)]{sakajo2019}
{\sc \au{Sakajo, Takashi}} \yr{2019}  \at{Exact solution to a liouville equation with stuart vortex distribution on the surface of a torus}.  \jt{Proceedings: Mathematical, Physical and Engineering Sciences}  \bvol{475}~(2224),  \pg{pp. 1--16}.

\bibitem[Smith \& Yakhot(1993)]{smith1993}
{\sc \au{Smith, Leslie~M.} \& \au{Yakhot, Victor}} \yr{1993}  \at{Bose condensation and small-scale structure generation in a random force driven 2d turbulence}.  \jt{Phys. Rev. Lett.}  \bvol{71},  \pg{352--355}.

\bibitem[Stieltjes(1900)]{stieltjes1900}
{\sc \au{Stieltjes, T.~J.}} \yr{1900}  \at{{Sur certains polynômes: Qui vérifient une équation différentielle linéaire du second ordre et sur la theorie des fonctions de Lamé}}.  \jt{Acta Mathematica}  \bvol{6}~(none),  \pg{321 -- 326}.

\bibitem[Stremler \& Aref(1999)]{stremler1999}
{\sc \au{Stremler, Mark~A.} \& \au{Aref, Hassan}} \yr{1999}  \at{Motion of three point vortices in a periodic parallelogram}.  \jt{Journal of Fluid Mechanics}  \bvol{392},  \pg{101–128}.

\bibitem[Stuart(1967)]{stuart1967}
{\sc \au{Stuart, J.~T.}} \yr{1967}  \at{On finite amplitude oscillations in laminar mixing layers}.  \jt{Journal of Fluid Mechanics}  \bvol{29}~(3),  \pg{417–440}.

\bibitem[Suri {\em et~al.\/}(2020)Suri, Kageorge, Grigoriev \& Schatz]{suri2020}
{\sc \au{Suri, Balachandra}, \au{Kageorge, Logan}, \au{Grigoriev, Roman~O.} \& \au{Schatz, Michael~F.}} \yr{2020}  \at{Capturing turbulent dynamics and statistics in experiments with unstable periodic orbits}.  \jt{Phys. Rev. Lett.}  \bvol{125},  \pg{064501}.

\bibitem[Suri {\em et~al.\/}(2017)Suri, Tithof, Grigoriev \& Schatz]{suri2017}
{\sc \au{Suri, Balachandra}, \au{Tithof, Jeffrey}, \au{Grigoriev, Roman~O} \& \au{Schatz, Michael~F}} \yr{2017}  \at{Forecasting fluid flows using the geometry of turbulence}.  \jt{Physical review letters}  \bvol{118}~(11),  \pg{114501}.

\bibitem[Suri {\em et~al.\/}(2018)Suri, Tithof, Grigoriev \& Schatz]{suri2018}
{\sc \au{Suri, Balachandra}, \au{Tithof, Jeffrey}, \au{Grigoriev, Roman~O} \& \au{Schatz, Michael~F}} \yr{2018}  \at{Unstable equilibria and invariant manifolds in quasi-two-dimensional kolmogorov-like flow}.  \jt{Physical Review E}  \bvol{98}~(2),  \pg{023105}.

\bibitem[Waleffe(2001)]{Waleffe2001}
{\sc \au{Waleffe, F.}} \yr{2001}  \at{Exact coherent structures in channel flow}.  \jt{Journal of Fluid Mechanics}  \bvol{435},  \pg{93–102}.

\bibitem[Waleffe(2003)]{waleffe2003}
{\sc \au{Waleffe, Fabian}} \yr{2003}  \at{Homotopy of exact coherent structures in plane shear flows}.  \jt{Physics of Fluids}  \bvol{15}~(6),  \pg{1517--1534}.

\bibitem[Wang {\em et~al.\/}(2007)Wang, Gibson \& Waleffe]{wang2007}
{\sc \au{Wang, Jue}, \au{Gibson, John} \& \au{Waleffe, Fabian}} \yr{2007}  \at{Lower branch coherent states in shear flows: Transition and control}.  \jt{Phys. Rev. Lett.}  \bvol{98},  \pg{204501}.

\bibitem[Wedin \& Kerswell(2004)]{wedin_kerswell_2004}
{\sc \au{Wedin, H.} \& \au{Kerswell, R.~R.}} \yr{2004}  \at{Exact coherent structures in pipe flow: travelling wave solutions}.  \jt{Journal of Fluid Mechanics}  \bvol{508},  \pg{333–371}.

\bibitem[Weiss \& McWilliams(1991)]{weiss1991}
{\sc \au{Weiss, Jeffrey~B.} \& \au{McWilliams, James~C.}} \yr{1991}  \at{{Nonergodicity of point vortices}}.  \jt{Physics of Fluids A: Fluid Dynamics}  \bvol{3}~(5),  \pg{835--844}.

\bibitem[Yaln\ifmmode\imath\else\i\fi{}z {\em et~al.\/}(2021)Yaln\ifmmode\imath\else\i\fi{}z, Hof \& Budanur]{yalniz2021}
{\sc \au{Yaln\ifmmode\imath\else\i\fi{}z, G\"okhan}, \au{Hof, Bj\"orn} \& \au{Budanur, Nazmi~Burak}} \yr{2021}  \at{Coarse graining the state space of a turbulent flow using periodic orbits}.  \jt{Phys. Rev. Lett.}  \bvol{126},  \pg{244502}.

\bibitem[Zhigunov \& Grigoriev(2023)]{Zhigunov2023}
{\sc \au{Zhigunov, Dmitriy} \& \au{Grigoriev, Roman~O.}} \yr{2023}  \at{Exact coherent structures in fully developed two-dimensional turbulence}.  \jt{Journal of Fluid Mechanics}  \bvol{970},  \pg{A18}.

\end{thebibliography}

\end{document}